\documentclass[prd,aps,showpacs,nofootinbib,tightenlines]{revtex4}
\usepackage{mathrsfs}
\usepackage{amsmath}
\usepackage{amssymb}
\usepackage{epsfig}
\usepackage{graphicx}
\usepackage{booktabs}
\usepackage{multirow}
\usepackage{subfigure}
\usepackage[section]{placeins}
\usepackage{float}
\usepackage{slashed}
\usepackage{longtable}
\usepackage[colorlinks, citecolor=blue, linkcolor=red, urlcolor=blue]{hyperref}

\allowdisplaybreaks[4]
\begin{document}
\makeatletter\def\Hy@Warning#1{}\makeatother
\newcommand{\psl}{ p \hspace{-1.8truemm}/ }
\newcommand{\nsl}{ n \hspace{-2.2truemm}/ }
\newcommand{\vsl}{ v \hspace{-2.2truemm}/ }
\newcommand{\epsl}{\epsilon \hspace{-1.8truemm}/\,  }
\title{ Semileptonic baryon decays $\Xi_b\rightarrow \Xi_c \ell^- \bar{\nu}_\ell $ in perturbative QCD}
\author{Zhou Rui$^1$ }\email{jindui1127@126.com}
\author{Zhi-Tian Zou$^1$ }\email{zouzt@ytu.edu.cn}
\author{Ya Li$^2$}\email{liyakelly@163.com}
\author{Ying Li$^1$ }\email{liying@ytu.edu.cn}
\affiliation{Department of Physics, Yantai University, Yantai 264005, China}
\affiliation{Department of Physics, College of Sciences, Nanjing Agricultural University,
Nanjing, Jiangsu 210095, China}
\date{\today}
\begin{abstract}
We perform a detailed analysis of the semileptonic $\Xi_b\rightarrow \Xi_c \ell^- \bar{\nu}_\ell$ decays within the perturbative QCD (PQCD) framework. In our study, the $\Xi_b\rightarrow \Xi_c$ transition form factors are calculated using several popular models for baryonic light-cone distribution amplitudes (LCDAs). These form factors are then employed to analyze a range of observable quantities for the semileptonic processes via the helicity formalism. Our work presents predictions for the branching fractions of these decays for both the $\tau$ and $e$ channels. Notably, the obtained lepton flavor universality ratio, $\mathcal{R}_{\Xi_c}\approx0.3$, may offer new insights into the $\mathcal{R}^{(*)}$ puzzle. Furthermore, we investigate various angular observables, such as forward-backward asymmetries, lepton-side convexity parameters, and polarization asymmetries, which provide complementary information regarding potential new physics in $b$-baryonic semileptonic transitions. The numerical results for these angular observables are presented as both functions of $q^2$ and as averaged values. We observe that the lepton mass plays a significant role in shaping the angular distributions, affecting most of the observables under consideration. These results are expected to be valuable for both current and future experimental investigations of semileptonic heavy-to-heavy baryon decays.


\end{abstract}



\maketitle

\section{Introduction}
Semileptonic bottom hadrons decays mediated via $b \rightarrow c$ charged-current interactions provide a powerful laboratory for testing lepton flavor universality (LFU) in the Standard Model (SM)~\cite{Bernlochner:2021vlv,Bifani:2018zmi}. In particular, the lepton universality ratios $\mathcal{R}_{H_c}$\footnote{$H_c$ denotes one of the charmed hadrons}, defined as the branching fraction ratios of  the semitauonic decays with respect to their muonic counterparts, hold promise for exploring new physics (NP) since both experimental and theoretical uncertainties can be brought under control. On the experimental front, several independent experiments, including BaBar, Belle, and LHCb, have found some lepton universality ratios deviate from their SM expectations. For example, a combination of measurements for $\mathcal{R}_D$ and $\mathcal{R}_{D^*}$~\cite{BaBar:2012obs, BaBar:2013mob, Belle:2016ure, Belle:2015qfa, Belle:2019rba, Belle:2017ilt, LHCb:2023uiv, LHCb:2023zxo, LHCb:2024jll} is in tension with evaluated in the SM~\cite{Bigi:2016mdz, Gambino:2019sif, Bordone:2019vic, Bernlochner:2017jka, Jaiswal:2017rve, Martinelli:2021onb, Bigi:2017jbd, FermilabLattice:2021cdg} at the level of three standard deviations ($\sigma$)~\cite{HFLAV:2022esi}, indicating a potential enhancement of the semitauonic transition rate. An analogous measurement using $B_c\rightarrow J/\psi $ semileptonic decays has been performed by LHCb~\cite{LHCb:2017vlu}. The yield value of  $\mathcal{R}_{J/\psi}$  again appears to be a little less than $2\sigma$  larger than its SM expectation \cite{Rui:2016opu, Qiao:2012vt, Harrison:2020nrv, Dutta:2017xmj}. These clear anomalies attract a lot of theory attention from the NP point of view; for reviews, see Refs.~\cite{Li:2018lxi,Pich:2019pzg,Iguro:2024hyk}.

In the $b$-baryon sector, LHCb recently reported the first observation of the $\Lambda_b\rightarrow \Lambda_c \ell^- \bar{\nu}_\ell $ decays~\cite{LHCb:2022piu} with a significance of $6.1\sigma$, where $\ell$ is a charged lepton. The experimental value of $\mathcal{R}_{\Lambda_c}$ seems to contradict the trend shown by the results of $\mathcal{R}_{D^{(*)}}$ and $\mathcal{R}_{J/\psi}$, which is slightly shifted toward a lower value than the predictions of SM~\cite{Bernlochner:2018bfn, Detmold:2015aaa, Bernlochner:2018kxh}. This opposite behavior is unexpected, as all these processes are described by the same effective Hamiltonian for $b\rightarrow c$ transition~\cite{Fedele:2022iib}.
In Refs.~\cite{Duan:2024ayo,Endo:2025fke},
some sum rules among these LFU ratios in  heavy quark symmetry were  proposed
to check consistency in the experimental results independently of new physics models.
Of course,  the measured value at the current precision is insufficient to demonstrate a substantial tension with the SM expectations. More precise measurements will be necessary to clarify this situation. On the other hand, it is important to check whether there are similar deviations between the SM predictions and experimental data in other $b$-baryon semileptonic decays or not. Any possible deviations will increase our hope to indirectly search for NP.

The weak semileptonic channel $\Xi_b\rightarrow \Xi_c \ell^- \bar{\nu}_\ell $ has not been observed yet, but the LHCb Collaboration recently observed the two-body modes, $\Xi^0_b\rightarrow \Xi_c^+ D_s^-$ and $\Xi^-_b\rightarrow \Xi_c^0 D_s^-$~\cite{LHCb:2023ngz}, suggests experimental measurement of $\Xi_b\rightarrow \Xi_c $ transition can be promising in the near future. Theoretically,  the $\Xi_b\rightarrow \Xi_c $ transition has been studied by various authors~\cite{Cheng:1996cs,Ebert:2006rp, Singleton:1990ye, Cheng:1995fe, Ivanov:1996fj, Ivanov:1998ya, Cardarelli:1998tq, Albertus:2004wj, Korner:1994nh,Dutta:2018zqp,Zhang:2019xdm,Ke:2024aux,Neishabouri:2025abl}. The predicted value $\mathcal{R}_{\Xi_c}$ within these models ranges from 0.255 to 0.34, which  can provide  us with valuable information. Studying such processes not only refines our understanding of determinations of the Cabibbo-Kobayashi-Maskawa (CKM) matrix element $V_{cb}$, but also reveals whether a similar anomaly also exists in $\mathcal{R}_{\Xi_c}$ and gives complimentary information regarding possible NP. However, a calculation of $\mathcal{R}_{\Xi_c}$ is not yet available from PQCD, and comparing it with future experimental data can open a new window for exploring LFU.

The present study focuses on the investigation of the  semileptonic decay of $\Xi_b\rightarrow \Xi_c \ell^- \bar{\nu}_\ell $ with $\ell=e,\mu,\tau$. We employ the PQCD approach based on the $k_T$ factorization~\cite{Keum:2000wi,Lu:2000em}, which is a well-established QCD-inspired approach in predicting heavy flavor hadronic weak decays~\cite{Kurimoto:2001zj,Lu:2002ny,Ali:2007ff,Wang:2010ni,Rui:2021kbn,Chai:2022ptk,Rui:2023fiz}. The light-cone distribution amplitudes are basic input ingredients to make predictions with the PQCD approach. In the last decade, the investigation of LCDAs of heavy mesons and baryons has made great progress~\cite{plb665197, jhep112013191, epjc732302, plb738334, jhep022016179, Ali:2012zza, Han:2024min, Han:2024yun, Wang:2024wwa}. However, the charmed baryon LCDAs  have received less attention in the literature, which limits the study of the decays containing charmed baryons. In this work, we will propose three simple models for the charmed baryon LCDAs based on the light-front formalism for baryons~\cite{Schlumpf:1992ce} and heavy-quark symmetry. With the universal nonperturbative LCDAs, we have    calculated the branching ratios, the lepton universality ratio, and other asymmetry observables of the concerned decays using the PQCD approach for the first time. Any future experimental data and their comparisons with the predictions of the present study will help us check whether there exists any discrepancy with the SM predictions in the channel under question or not.

The reminder of this paper is organized as follows: In Section~\ref{sec:framework}, we start with kinematics and the relevant LCDAs responsible for the transitions under consideration. A brief discussion on $\Xi_b \to \Xi_c$ transition form factors and  helicity amplitudes is also presented. According to the angular distribution, we also write down several observables, such as the differential branching ratios, the forward-backward asymmetries, the longitudinal and transverse polarizations of lepton and $\Xi_c$ baryon, and the convexity parameters for the $\Xi_b \to \Xi_c\ell^- \bar{\nu}_\ell $ decays. In section~\ref{sec:results}, we conduct a numerical analysis of the baryonic form factors by determining the working regions of auxiliary parameters and find the fit functions for the behavior of form factors in terms of transferred momentum squared. We determine  the differential and integrated observables  for all the lepton channels and compare our results with the predictions of other theoretical studies. We reserve the last section for the summary and conclusions. Some details of the calculations are presented in the appendix.

\section{Formalism}\label{sec:framework}
\subsection{Kinematics and baryonic light cone distribution amplitudes }
\begin{figure}[!htbh]
	\begin{center}
	   \vspace{4cm}  \centerline{\epsfxsize=8cm \epsffile{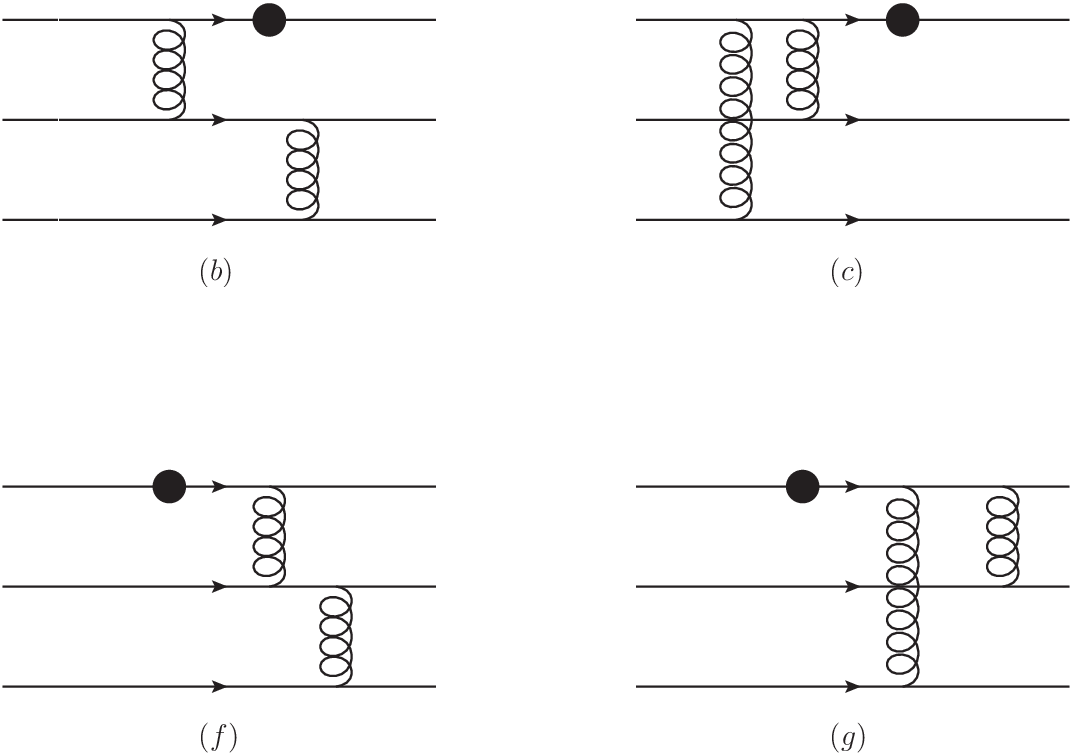}}
		\caption{
Lowest-order diagrams for $\Xi_b\rightarrow \Xi_c$ transition form factors. }
		\label{fig:C}
	\end{center}
\end{figure}

The $\Xi_b\rightarrow \Xi_c \ell^- \bar{\nu}_\ell $ decay channel proceeds via $b \rightarrow c \ell^- \bar{\nu}_\ell$ at quark level. The corresponding Feynman diagrams at the leading order are displayed in Fig.~\ref{fig:C}, where two hard gluons attach the three incoming and outgoing quarks in all possible ways.
Note that, here, we only display one of the diagrams that are associated by swapping two light quarks, whose amplitudes are equal because of the symmetry of the baryonic wave functions~\cite{Zhang:2022iun}.
For simplicity, we work in the rest frame of the $\Xi_b$ baryon and use light cone coordinates. The momentum of the $\Xi_b$ and $\Xi_c$ can be denoted as
\begin{eqnarray}
p=\frac{M}{\sqrt{2}}\left(1,1,\textbf{0}_{T}\right), \quad p'=\frac{M}{\sqrt{2}}\left(f^+,f^-,\textbf{0}_{T}\right),
\end{eqnarray}
where $M$ is the mass of the  $\Xi_b$ baryon. The factors $f^{\pm}=(f\pm \sqrt{f^2-1})r$ are defined in terms of the velocity transfer, $f=\frac{p \cdot  p'}{M^2r}$ with $r=m/M$ being the mass ratio between  $\Xi_c$ and $\Xi_b$  baryons. It is easy to obtain the momentum transfer, $q=p-p'$,  satisfies $f=\frac{(1+r^2)M^2-q^2}{2rM^2}$. The valence quark moments are parametrized as
\begin{eqnarray}
 k_{2}=\left(0,\frac{M}{\sqrt{2}}x_{2},\textbf{k}_{2T}\right), \quad  k_{3}=\left(0,\frac{M}{\sqrt{2}}x_{3},\textbf{k}_{3T}\right), \quad k_1=p-k_2-k_3,
\end{eqnarray}
for $\Xi_b$ baryon  and
\begin{eqnarray}
k'_{2}=\left(\frac{M}{\sqrt{2}}f^+x'_{2},0,\textbf{k}'_{2T}\right), \quad  k'_{3}=\left(\frac{M}{\sqrt{2}}f^+x'_{3},0,\textbf{k}'_{3T}\right), \quad k'_1=p'-k'_2-k'_3,
\end{eqnarray}
for $\Xi_c$ baryon. Here $k^{(')}_1$ is associated with the heavy $b(c)$ quark. $\textbf{k}^{(')}_{iT}$ and $x^{(')}_i$ with $i=1,2,3$ represent the transverse momentum and longitudinal momentum fractions of the valence quarks inside the baryons, respectively. They satisfy the momentum conservation conditions:
 \begin{eqnarray}
  \sum_{l=1}^3x^{(')}_l=1,\quad \sum_{l=1}^3\textbf{k}^{(')}_{lT}=0.
 \end{eqnarray}
Similar to the mesonic $B\rightarrow D$ transition~\cite{Kurimoto:2002sb}, the $\Xi_b\rightarrow \Xi_c $  transition also involves multiple scales, such as the heavy baryon masses $M$ and $m$ and the heavy quark masses $m_b$ and $m_c$. To simplify the formalism, we shall neglect the mass differences between heavy baryons and heavy quarks and make the assumptions $ m_b\approx M$ and $m_c\approx m$~\cite{Shih:1999yh}. This approximation ensures that the $\Xi_b\rightarrow \Xi_c $ transition form factors are real, which is also required by time reversal invariance~\cite{Schlumpf:1992ce}. It also indicates that the momentum fractions of heavy quarks are at the order $\mathcal{O}(m^2_b/M^2) \sim \mathcal{O}(m^2_c/m^2)\sim 1$. This character is subject to strong  kinematic constraints  on the baryonic LCDAs, which will be discussed in detail later.

\begin{figure}[!htbh]
\begin{center}
\setlength{\abovecaptionskip}{0pt}
\centerline{
\hspace{1cm}\subfigure{\epsfxsize=4.5cm \epsffile{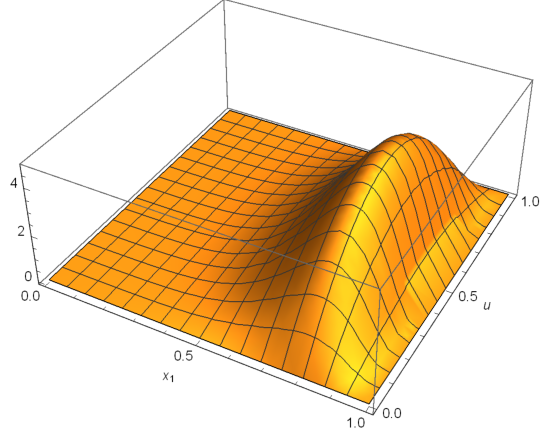} }
\hspace{1cm}\subfigure{ \epsfxsize=4.5cm \epsffile{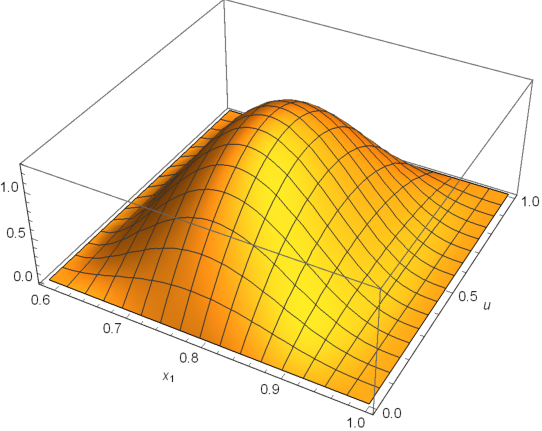}}
\hspace{1cm}\subfigure{ \epsfxsize=4.5cm \epsffile{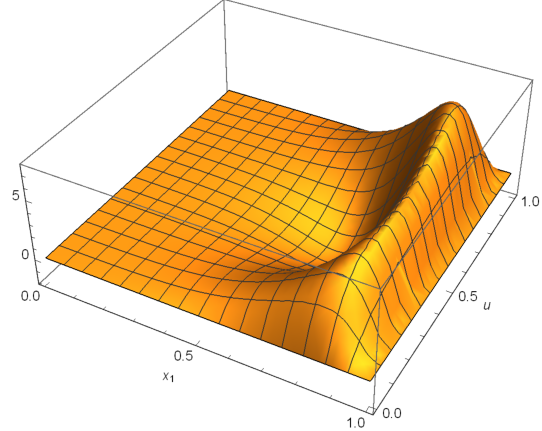}}}
\vspace{1cm}
\caption{The twist 2 LCDAs of  $\Xi_b$ for the Exponential model (left), QCDSR model (middle), and Gegenbauer model (right). }
 \label{fig:xib}
\end{center}
\end{figure}

In the course of the PQCD calculations, the necessary inputs contain the LCDAs, which are constructed via the nonlocal matrix elements. In the heavy-quark effective theory (HQET), baryons with one heavy quark $Q$ ($b$ or $c$) can be regarded as  a quark-diquark system, where the heavy quark spin decouples from the light diquark dynamics in the leading order of the heavy quark mass expansion. In this simplified picture, the ground-state baryons  with the spin parity $J^P$ are determined by the spin parity $j^p$ of the diquark. For the SU(3) flavor antitriplet, the three-quark  LCDAs can be described by a set of two-particle LCDAs corresponding to the diquark state. The corresponding matrix element can be parametrized as~\cite{plb665197,jhep112013191,epjc732302,plb738334,jhep022016179,Ali:2012zza}
\begin{eqnarray}\label{eq:LCDAs20}
\epsilon^{ijk}\langle0|q^i_{1\alpha}(t_1)q^j_{2\beta}(t_2)Q^k_\gamma(0)|\Xi_Q \rangle &=&\frac{ f^{(1)}}{8}
[(\slashed{\bar n}\gamma_5C)_{\alpha\beta}\widetilde{\phi}_2^Q(t_1,t_2)+(\slashed{ n}\gamma_5C)_{\alpha\beta}\widetilde{\phi}_4^Q(t_1,t_2)]u_\gamma \nonumber\\
&&+\frac{f^{(2)}}{4}[(\gamma_5C)_{\alpha\beta}\widetilde{\phi}_{3s}^Q(t_1,t_2)-\frac{i}{2}(\sigma_{\bar nn}\gamma_5C)_{\alpha\beta}\widetilde{\phi}^{Q}_{3a}(t_1,t_2)]u_\gamma
\end{eqnarray}
where $\sigma_{\bar nn}=\frac{i}{2}(\slashed{\bar n}\slashed{n}-\slashed{ n}\slashed{\bar n})$ with $n,\bar n$ being two light-cone vectors satisfy $n^2=\bar n^2=0$ and  $n \cdot \bar n=2$. The color indices are denoted by $i$, $j$, and $k$  and $\epsilon^{ijk}$ is a totally antisymmetric tensor. $C$ denotes the charge conjugation matrix, and the subscripts $\alpha,\beta,\gamma$ are the spinor indices. $u_\gamma$ denotes the heavy quark spinor, and its momentum dependence and spin index are suppressed in the notation. The values $f^{(1,2)}$ are taken as  $f^{(1,2)}_{\Xi_b}=0.032\pm 0.009$ GeV$^3$ and  $f^{(1,2)}_{\Xi_c}=0.027\pm 0.008$ GeV$^3$ from the QCD sum rules~\cite{Wang:2010fq}. The symbols $\widetilde{\phi}_{i}^Q$ with $i=2,4,3s,3a$ represent four different LCDAs, and the numbers in the subscript indicate the twist. They have definite symmetry properties in the limit of the exact SU(3) flavor symmetry, such as $\widetilde{\phi}_{3a}^Q$ being antisymmetric under the exchange $t_1\leftrightarrow t_2$, while others are symmetric. They are normalized as  $\widetilde{\phi}^Q_{3a}(0,0)=0$ and $\widetilde{\phi}^Q_{2}(0,0)=\widetilde{\phi}^Q_{4}(0,0)=\widetilde{\phi}^Q_{3s}(0,0)=1$, respectively. Going over to momentum space, we define~\cite{plb665197}
 \begin{eqnarray}
\widetilde{\phi}^Q_i(t_1,t_2)=\int_0^\infty\omega d\omega \int_0^1 du \phi^Q_i(\omega,u)e^{-i\omega[t_1u+t_2(1-u)]},
\end{eqnarray}
where $\omega=(x_2+x_3)M$ is the total energy carried by the two light quarks in the heavy-quark rest frame. $u=x_2/(x_2+x_3)$ is the energy fraction carried by the lighter quark in the diquark system. For the $\Xi_b$ baryon, four popular models of $\phi^b_i(\omega,u)$ up to twist 4 exist in literature\footnote{Note that the Exponential,  QCDSR, and Free parton models are only used to study the LCDAs of $\Lambda_b$ baryons.  Since $\Lambda_b$ and $\Xi_b$ belong to the same antitriplet,  it is reasonable to believe that these models can describe the $\Xi_b$ one as well.}:
\begin{itemize}
  \item Exponential model~\cite{jhep112013191}:
  \begin{eqnarray}\label{eq:exx}
\phi^b_2(\omega,u)&=&    \frac{\omega^2u(1-u)}{\omega_0^4}e^{-\frac{\omega}{\omega_0}},\nonumber\\
\phi^b_{3s}(\omega,u)&=&\frac{\omega}  {2\omega_0^3}e^{-\frac{\omega}{\omega_0}},\nonumber\\
\phi^b_{3a}(\omega,u)&=&\frac{\omega(2u-1)}  {2\omega_0^3}e^{-\frac{\omega}{\omega_0}},\nonumber\\
\phi^b_4(\omega,u)&=&     \frac{1}     {\omega_0^2}e^{-\frac{\omega}{\omega_0}},
\end{eqnarray}
where $\omega_0=0.4\pm0.1$ GeV measures the average of the two light quarks inside the $\Xi_b$ baryon.
  \item QCDSR model~\cite{plb665197}:
  \begin{eqnarray}\label{eq:qcd}
\phi^b_2    (\omega,u)&=&     \frac{15}{2\mathcal{N}}\omega^2 u (1-u) \int_{\frac{\omega}{2}}^{s_0}dse^{-s/\tau}(s-\frac{\omega}{2}),\nonumber\\
\phi^b_{3s}(\omega,u)&=&     \frac{15}{4\mathcal{N}}\omega           \int_{\frac{\omega}{2}}^{s_0}dse^{-s/\tau}(s-\frac{\omega}{2})^2,\nonumber\\
\phi^b_{3a}(\omega,u)&=&     \frac{15}{4\mathcal{N}}\omega  (2u-1)   \int_{\frac{\omega}{2}}^{s_0}dse^{-s/\tau}(s-\frac{\omega}{2})^2,\nonumber\\
\phi^b_4    (\omega,u)&=&     \frac{5}{\mathcal{N}}                   \int_{\frac{\omega}{2}}^{s_0}dse^{-s/\tau}(s-\frac{\omega}{2})^3,
\end{eqnarray}
where $\mathcal{N}=\int_0^{s_0}dse^{-s/\tau}s^5$ with the continuum threshold $s_0=1.2$ GeV and the Borel parameter $\tau=0.6\pm 0.2$ GeV. Note that the variable $\omega$ is restricted to a range of $0<\omega< 2s_0$  when performing a integration~\cite{plb665197}.
 \item Gegenbauer  model~\cite{epjc732302}:
  \begin{eqnarray}\label{eq:g2}
\phi^b_2  (\omega,u)&=&   \omega^2u(1-u) \sum_{l=0}^2\frac{a_l}{\epsilon_l^4}\frac{c_l^{3/2}(2u-1)}{|c_l^{3/2}|^2}e^{-\frac{\omega}{\epsilon_l}},\nonumber\\
\phi^b_{3s,3a}(\omega,u)&=& \frac{\omega}{2}  \sum_{l=0}^2\frac{a_l}{\epsilon_l^3}\frac{c_l^{1/2}(2u-1)}{|c_l^{1/2}|^2}e^{-\frac{\omega}{\epsilon_l}},\nonumber\\
\phi^b_4  (\omega,u)&=&   \sum_{l=0}^2\frac{a_l}{\epsilon_l^2}\frac{c_l^{1/2}(2u-1)}{|c_l^{1/2}|^2}e^{-\frac{\omega}{\epsilon_l}},
\end{eqnarray}
where the shape parameters $a_l$ and $\epsilon_l$ dependence on a free parameter, $A=0.5\pm 0.2$, whose definite forms can be found  in~\cite{epjc732302}.
\item Free parton model~\cite{jhep112013191}:
   \begin{eqnarray}\label{eq:fp}
\phi^b_2    (\omega,u)&=&     \frac{15\omega^2u(1-u)(2\bar{\Lambda}-\omega)  }{4\bar{\Lambda}^5}\Theta(2\bar{\Lambda}-\omega),\nonumber\\
\phi^b_{3s}(\omega,u)&=&     \frac{15\omega(2\bar{\Lambda}-\omega)^2  }{16\bar{\Lambda}^5}\Theta(2\bar{\Lambda}-\omega),\nonumber\\
\phi^b_{3a}(\omega,u)&=&     \frac{15\omega(2u-1) (2\bar{\Lambda}-\omega)^2  }{16\bar{\Lambda}^5}\Theta(2\bar{\Lambda}-\omega),\nonumber\\
\phi^b_4    (\omega,u)&=&     \frac{5 (2\bar{\Lambda}-\omega)^3  }{8\bar{\Lambda}^5}\Theta(2\bar{\Lambda}-\omega),
\end{eqnarray}
where $2\bar{\Lambda}=M-m_b$ and $\Theta(x)$ denotes a step function.
\end{itemize}
It is worth emphasizing that the parameter $\bar{\Lambda}$ in the Free parton model vanishes under the assumption $M\approx m_b$ as previously stated. Hence, we will discard the Free parton model and focus on the other three models for the numerical analysis and discussion hereafter. The qualitative behaviors of the twist-2 LCDAs  for the Exponential model, QCDSR model, and Gegenbauer model are visualized in Fig.~\ref{fig:xib}. Here $x_1=1-\omega/M$ is the $b$ quark momentum fraction inside $\Xi_b$ baryon, and $x_1$-dependence indicates the dynamic distributions of the $b$ quark and the light diquark. It can be seen that the maximums of the distribution for all the three models are shifted towards the $b$ quark,  such that the $b$ quark momentum squared $k^2_1$ is roughly equal to $M^2$. This behavior is consistent with our previous assumption of neglecting the mass difference between heavy quark and  heavy baryon. The $u$-dependence denotes dynamic distributions of the light quarks in the diquark system. We observe that the curves  of both the Exponential model and QCDSR model are symmetric under the interchange $u\leftrightarrow 1-u$ and the axis of symmetry is located at $u=0.5$. However, this symmetry is broken  for the Gegenbauer model since the SU(3) breaking correction is taken into account. It means that the position of the peak will deviate $u=0.5$ because the $s$ quark is heavier, as shown in the last diagram of Fig.~\ref{fig:xib}.

\begin{figure}[!htbh]
\begin{center}
\setlength{\abovecaptionskip}{0pt}
\centerline{
\hspace{1cm}\subfigure{\epsfxsize=4.5cm \epsffile{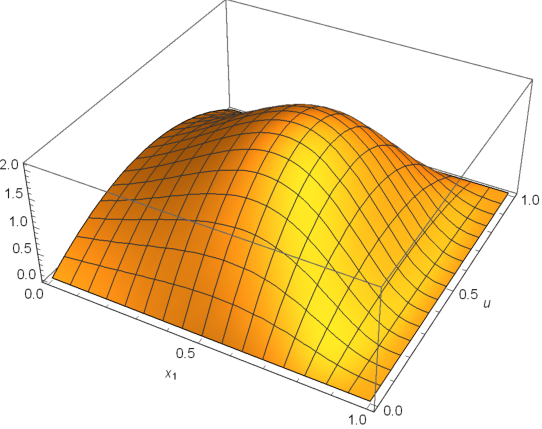} }
\hspace{1cm}\subfigure{ \epsfxsize=4.5cm \epsffile{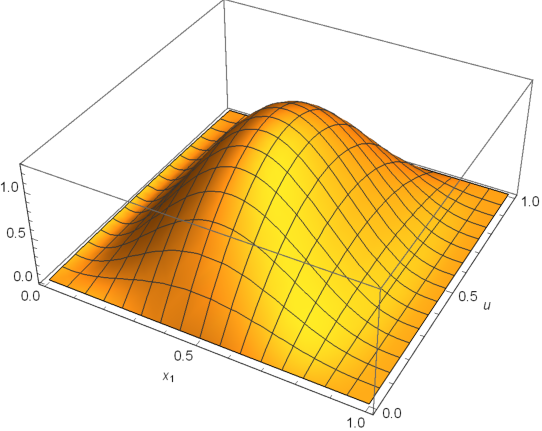}}
\hspace{1cm}\subfigure{ \epsfxsize=4.5cm \epsffile{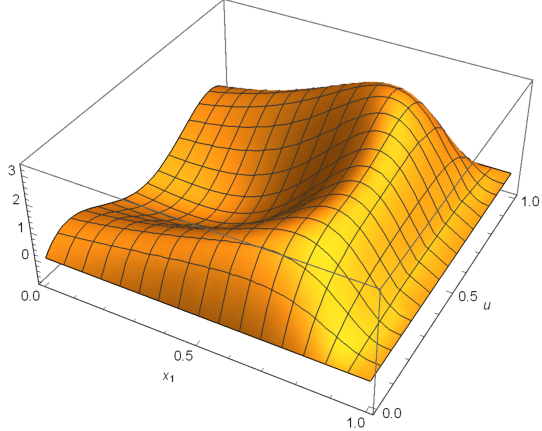}}}
\vspace{1cm}
\caption{Same as Fig.~\ref{fig:xib}  but for the LCDAs of $\Xi_c$.}
 \label{fig:xic}
\end{center}
\end{figure}

\begin{figure}[!htbh]
\begin{center}
\setlength{\abovecaptionskip}{0pt}
\centerline{
\hspace{1cm}\subfigure{\epsfxsize=4.5cm \epsffile{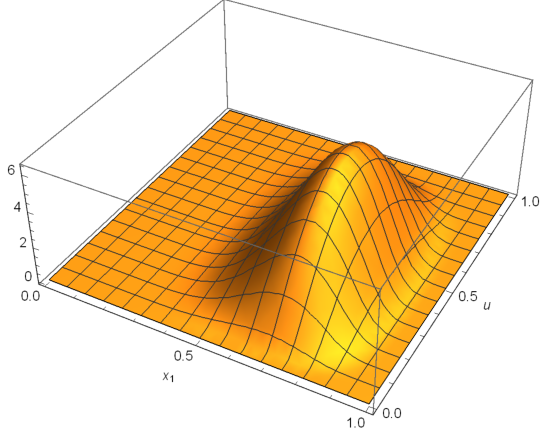} }
\hspace{1cm}\subfigure{ \epsfxsize=4.5cm \epsffile{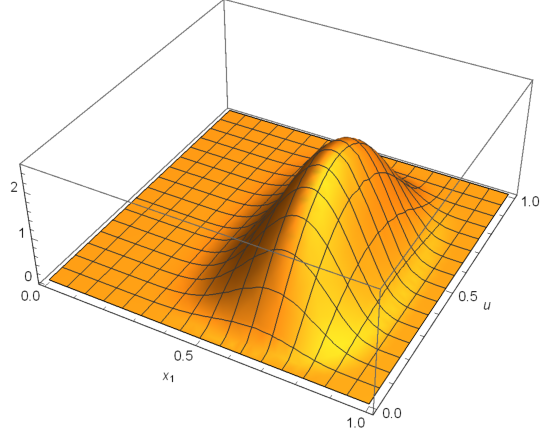}}
\hspace{1cm}\subfigure{ \epsfxsize=4.5cm \epsffile{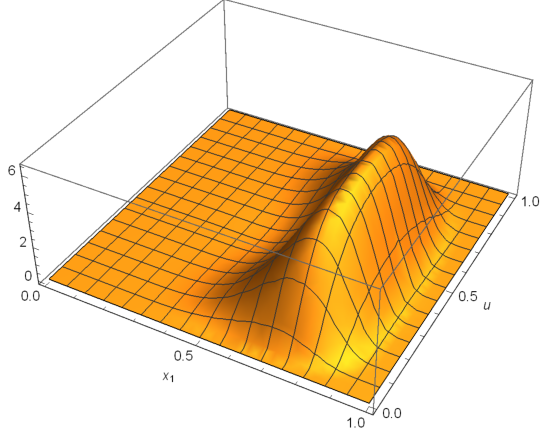}}}
\vspace{1cm}
\caption{The modified twist 2 LCDAs of  $\Xi_c$ for the Exponential model (left), QCDSR model (middle), and Gegenbauer model (right). }
 \label{fig:xicp}
\end{center}
\end{figure}

The LCDAs of $\Xi_c$ baryon are still poorly known at the moment. Based on heavy-quark symmetry, we can use the same forms and parameters as the LCDAs of $\Xi_b$ with obvious mass replacement,
\begin{eqnarray}
\phi^c_i(\omega,u)=\phi^b_i(\omega,u)|_{M\rightarrow m}.
\end{eqnarray}
This results in the $\Xi_c$ baryon LCDAs behave differently as displayed in Fig.~\ref{fig:xic}. These functions for $\Xi_c$ decreases more slowly in the small $x_1$ region than those for the $\Xi_b$ ones. Moreover, it is  unexpected that  the peaks for the Exponential model and QCDSR model  are located at smaller $x_1(\sim 0.5)$ with broader width. To ensure that the valence charm quark is close to the mass shell, following the practice in the literature~\cite{Shih:1999yh, Schlumpf:1992ce}, we modify the $\Xi_c$ baryon LCDAs by multiplying by an exponential factor
\begin{eqnarray}\label{eq:ex}
 \phi^c_i(\omega,u)\rightarrow  N_i\phi^c_i(\omega,u) e^{-\frac{m^2}{2\beta^2x_1}-\frac{m_q^2}{2\beta^2x_2}-\frac{m_q^2}{2\beta^2x_3}}
\end{eqnarray}
with $N_i$ being the normalization constants. Several model parameters, such as the constituent quark masses $m_{u,d,s}$ and the confinement scale parameter $\beta$, are introduced. Presently, we do not have enough experimental information to determine them unambiguously. Here, we comply with the isospin symmetry and use the values $m_u=m_d=0.22$ GeV and $m_s=0.419$ GeV from the semirelativistic potential model~\cite{Li:2021qod}. Notice that the  SU(3)-breaking effect is reflected in this factor because $m_s>m_{u,d}$.  The shape parameter $\beta$ is adopted as $1.0\pm 0.2$ GeV according to Refs.~\cite{Shih:1999yh,prd59094014} to make the $\Lambda_b\rightarrow p,\Lambda_c$ form factors are dominated by the perturbative contributions. The plots of the modified twist-2 LCDAs of the $\Xi_c$ in the $x_1$ and $u$ distributions are presented in Fig.~\ref{fig:xicp}. It is obvious that the falloff for all three functions becomes strong in the low $x_1$ region, and the behaviors in the high $x_1$ region are closer to the counterparts in Fig.~\ref{fig:xib}. This improvement also increases peak positions to around $x_1\sim 0.7$, such that the charm quark, carrying the momentum squared $k'^2_1\approx m^2$, is almost on shell. Furthermore,  the modified LCDAs  with three different models show similar $x_1$ and $u$ distributions, which greatly reduce the model dependence of our predictions, as can be seen in the numerical computations. For convenience, in the following numerical calculations, we assume that the exponential factor in Eq.~(\ref{eq:ex}) is universal for all other higher-twists LCDAs of $\Xi_c$.

\subsection{ The baryonic form factors and hadronic helicity amplitudes}
The amplitude for the semileptonic decay $\Xi_b\rightarrow \Xi_c \ell^- \bar{\nu}_\ell $ can be written as
\begin{eqnarray}
\mathcal{M}=\frac{G_F}{\sqrt{2}}V_{cb} \langle\Xi_c(p')| \bar c\gamma_\mu (1-\gamma_5)b|\Xi_b(p)\rangle \bar \ell \gamma^\mu (1-\gamma_5)\nu_\ell,
\end{eqnarray}
where $G_F$ is the Fermi coupling constant and $V_{cb}$ is one of the elements of the CKM matrix. The hadronic matrix elements of the vector and axial-vector currents, containing all QCD dynamics, can be parametrized in terms of six invariant form factors as
\begin{eqnarray}\label{eq:FFs}
M_\mu^V&=&\langle \Xi_c|\bar {c}\gamma_\mu b |\Xi_b\rangle=\bar{u}_{\Xi_c}(p',\lambda') \left[\gamma_\mu f_1(q^2)- i\sigma_{\mu\nu}\frac{q^\nu}{M}f_2(q^2)+\frac{q_\mu}{M}f_3(q^2)\right]u_{\Xi_b}(p,\lambda),\nonumber\\
M_\mu^A&=&\langle \Xi_c|\bar {c}\gamma_\mu\gamma_5 b |\Xi_b\rangle=\bar{u}_{\Xi_c}(p',\lambda') \left[\gamma_\mu g_1(q^2)- i\sigma_{\mu\nu}\frac{q^\nu}{M}g_2(q^2)+\frac{q_\mu}{M}g_3(q^2)\right]\gamma_5u_{\Xi_b}(p,\lambda),
\end{eqnarray}
where $\sigma_{\mu\nu}=\frac{i}{2}[\gamma_\mu,\gamma_\nu]$. $\bar{u}_{\Xi_c}(p',\lambda')$  and $u_{\Xi_b}(p,\lambda)$ are Dirac spinors of the $\Xi_c$ and $\Xi_b$ baryons, and the involved variables $p^{(')}$  and $ \lambda^{(')}$ are the corresponding momentum and helicity, respectively. $f_i$ and $g_i$ with $i=1,2,3$ are three form factors describing the vector and axial transitions, respectively. Another popular parametrization considering the Lorentz invariance and parity reads,
\begin{eqnarray}
M_\mu^V&=&\bar{u}_{\Xi_c}(p',\lambda') \left[\gamma_\mu F_1(q^2)+F_2(q^2)\frac{p_\mu}{M}+F_3(q^2)\frac{p'_\mu}{M}\right]u_{\Xi_b}(p,\lambda),\nonumber\\
M_\mu^A&=&\bar{u}_{\Xi_c}(p',\lambda') \left[\gamma_\mu G_1(q^2)+G_2(q^2)\frac{p_\mu}{M}+G_3(q^2)\frac{p'_\mu}{M}\right]\gamma_5u_{\Xi_b}(p,\lambda),
\end{eqnarray}
which are convenient to compute in PQCD. The pertinent relation between the two sets of form factors can be expressed as follows:
\begin{eqnarray}
f_1&=&F_1+\frac{1}{2}(F_2+F_3)(1+r), \quad f_2=-\frac{1}{2}(F_2+F_3), \quad f_3=\frac{1}{2}(F_2-F_3),\nonumber\\
g_1&=&G_1-\frac{1}{2}(G_2+G_3)(1-r), \quad g_2=-\frac{1}{2}(G_2+G_3), \quad g_3=\frac{1}{2}(G_2-G_3).
\end{eqnarray}

The factorization formula  $F_i/G_i$ in the PQCD formalism can be written as
\begin{eqnarray}\label{eq:FG}
F_i/G_i=\frac{64f_{\Xi_b} f_{\Xi_c}  \pi^2 G_F}{27}\sum_k
\int\mathcal{D}x\mathcal{D}b
\alpha_s^2(t_k)e^{-S_{\Xi_b}-S_{\Xi_c}}\Omega_k(b_i,b'_i) H^{F_i/G_i}_k(x_i,x'_i),
\end{eqnarray}
where $k$ runs from $a$ to $g$ as shown in Fig.~\ref{fig:C}. The integration measures are given by
\begin{eqnarray}
\mathcal{D}x&=&dx_1dx_2dx_3\delta(1-x_1-x_2-x_3)dx'_1dx'_2dx'_3\delta(1-x'_1-x'_2-x'_3),\nonumber\\
\mathcal{D}b&=& d^2\textbf{b}_2d^2\textbf{b}_3d^2\textbf{b}'_2d^2\textbf{b}'_3.
\end{eqnarray}
The $\delta$ functions enforce momentum conservation. $H_{k}$ is the numerator of the hard amplitude depending on the spin structure of final state. $\Omega_{k}$ is the Fourier transformation of the denominator of the hard amplitude from the $k_T$ space to its conjugate $b$ space. The involving variables are $b^{(')}_i$  conjugate to the parton transverse momenta $k^{(')}_{iT}$. The hard scale $t_k$ for each diagram is chosen as the maximal virtuality of internal particles, including the factorization scales in a hard amplitude:
\begin{eqnarray}\label{eq:ttt}
t_k=\max(\sqrt{|t_A|},\sqrt{|t_B|},\sqrt{|t_C|},\sqrt{|t_D|},w,w'),
\end{eqnarray}
where $w^{(')}$ is chosen to be the smallest inverse of a typical transverse distance among three valence quarks in baryon~\cite{prd59094014}. $t_{A,B}$ are relevant to the two hard gluons, while $t_{C,D}$ are associated with the two virtual quarks. Those quantities associated with specific diagram, such as  $H_{k}$, and $t_{k}$, are collected in Appendix.

The Sudakov exponents associated with  the initial and final states in Eq.~(\ref{eq:FG}) are written as~\cite{Ali:2007ff,prd80034011}
\begin{eqnarray}\label{eq:sud}
S_{\Xi_b}&=&\sum_{l=2}^3s(k_l^{-},w)+\frac{8}{3}\int^t_{w}d \bar \mu \frac{\gamma(\alpha_s(\bar \mu))}{\bar \mu},\nonumber\\
S_{\Xi_c}&=&s_c(k_1^{'+},cw')+\sum_{l=2}^3s(k_l^{'+},cw')+\frac{8}{3}\int^t_{cw'}d \bar \mu \frac{\gamma(\alpha_s(\bar \mu))}{\bar \mu},
\end{eqnarray}
with the quark anomalous dimension $\gamma(\alpha_s(\bar \mu))=-\alpha_s(\bar \mu)/\pi$. The integral containing $\gamma(\alpha_s(\bar \mu))$, corresponding to the single logarithm RG evolution, describes the evolution from $t$ to the factorization scales. The functions $s$ and $s_c$ correspond to the double-logarithm evolution for an energetic light quark and charm quark, respectively. Their expressions at the next-to-leading-logarithm accuracy can be found in Refs.~\cite{prd59094014, Ali:2007ff, Liu:2023kxr, Liu:2020upy}. For consistency, we also adopt a two-loop running coupling constant and the involved scale $\Lambda_{\text{QCD}}=0.225$ GeV for the flavors $n_f=5$, which is determined by the experimental value of $\alpha_s(M_Z)=0.1179$~\cite{pdg2024}. The parameter $c$ is introduced to mimic theoretical uncertainties in resummation, whose values had been determined as $c=1.05$ for the charmed baryon~\cite{Rui:2024xgc}.

The relation between the hadronic matrix elements and the helicity amplitudes is defined as~\cite{Korner:1989qb,Dutta:2018zqp,Ray:2018hrx}
\begin{eqnarray}
H^{V/A}_{\lambda'\lambda_W}=M_\mu^{V/A}(\lambda')\epsilon^{\ast \mu}(\lambda_W),
\end{eqnarray}
where $\lambda'$ and $\lambda_W$ denote the respective helicities of the daughter baryon and  off-shell $W$ boson. Angular momentum conservation fixes the helicity $\lambda$ of the parent baryon such that $\lambda=\lambda'-\lambda_W$ \footnote{Here, the $z$-axis is chosen in the $\Xi_b$  rest frame such that the $\Xi_c$ travels in positive $z$-direction and the $W$ boson in negative $z$-direction.}.  $\epsilon^{\mu}$ is the polarization vectors of the virtual $W$ boson. There are four helicities for the off-shell $W$ boson, namely $\lambda_W=0,\pm1 $  for the vector case and $\lambda_W=0 $ for the scalar case. To distinguish the two $\lambda_W=0$ states, the scalar one, which does not contribute to the semileptonic decay in the lepton massless limit, is usually written as $\lambda_W=t$ with $t$ meaning temporal. Since we will also discuss lepton mass effects, it is necessary to include the scalar form factors $f_3$ and $g_3$. After lengthy calculations according to the standard prescriptions in Refs.~\cite{Gutsche:2013pp, Datta:2017aue, Kadeer:2005aq}, we obtain the helicity amplitudes, which are expressed through the decay form factors by the following relations:
\begin{eqnarray}
H^V_{\frac{1}{2}t}&=&\sqrt{\frac{Q_+^2-q^2}{q^2}}[Q_-f_1+\frac{q^2}{M}f_3],\nonumber\\
H^A_{\frac{1}{2}t}&=&\sqrt{\frac{Q_-^2-q^2}{q^2}}[Q_+g_1-\frac{q^2}{M}g_3],\nonumber\\
H^V_{\frac{1}{2}0}&=&\sqrt{\frac{Q_-^2-q^2}{q^2}}[Q_+f_1+\frac{q^2}{M}f_2],\nonumber\\
H^A_{\frac{1}{2}0}&=&\sqrt{\frac{Q_+^2-q^2}{q^2}}[Q_-g_1-\frac{q^2}{M}g_2],\nonumber\\
H^V_{\frac{1}{2}1}&=&\sqrt{2(Q_-^2-q^2)}[-f_1-\frac{Q_+}{M}f_2],\nonumber\\
H^A_{\frac{1}{2}1}&=&\sqrt{2(Q_+^2-q^2)}[-g_1+\frac{Q_-}{M}g_2],
\end{eqnarray}
where $Q_{\pm}=M\pm m$. The helicity-flipped amplitudes can be derived using the relations $H^V_{-\lambda'-\lambda_W}=H^V_{\lambda'\lambda_W}$ and $H^A_{-\lambda'-\lambda_W}=-H^A_{\lambda'\lambda_W}$~\cite{Faustov:2016pal}. The total amplitude is thus expressed by $H_{\lambda'\lambda_W}=H^V_{\lambda'\lambda_W}-H^A_{\lambda'\lambda_W}$ for the $V-A$ current.

For the convenience of describing the angular distribution, we can define some interesting  helicity structure functions with definite parity properties  in terms of bilinear combinations of helicity amplitudes. The relevant parity-conserving helicity structures are expressed as~\cite{Gutsche:2015mxa}
\begin{eqnarray}\label{eq:H1}
H_T(q^2)&=&|H_{\frac{1}{2}1}|^2+|H_{-\frac{1}{2}-1}|^2,\nonumber\\
H_L(q^2)&=&|H_{\frac{1}{2}0}|^2+|H_{-\frac{1}{2}0}|^2,\nonumber\\
H_S(q^2)&=&|H_{\frac{1}{2}t}|^2+|H_{-\frac{1}{2}t}|^2,\nonumber\\
H_{SL}(q^2)&=&Re[H_{\frac{1}{2}0}H^{\dag}_{\frac{1}{2}t}+H_{-\frac{1}{2}0}H^{\dag}_{-\frac{1}{2}t}],\nonumber\\
H_{ST}(q^2)&=&Re[H_{\frac{1}{2}1}H^{\dag}_{-\frac{1}{2}t}+H_{-\frac{1}{2}-1}H^{\dag}_{\frac{1}{2}t}],\nonumber\\
H_{LT}(q^2)&=&Re[H_{\frac{1}{2}1}H^{\dag}_{-\frac{1}{2}0}+H_{-\frac{1}{2}-1}H^{\dag}_{\frac{1}{2}0}],
\end{eqnarray}
and the parity-violating helicity structures as
\begin{eqnarray}\label{eq:H2}
H_{TP}(q^2)&=&|H_{\frac{1}{2}1}|^2-|H_{-\frac{1}{2}-1}|^2,\nonumber\\
H_{LP}(q^2)&=&|H_{\frac{1}{2}0}|^2-|H_{-\frac{1}{2}0}|^2,\nonumber\\
H_{SP}(q^2)&=&|H_{\frac{1}{2}t}|^2-|H_{-\frac{1}{2}t}|^2,\nonumber\\
H_{SLP}(q^2)&=&Re[H_{\frac{1}{2}0}H^{\dag}_{\frac{1}{2}t}-H_{-\frac{1}{2}0}H^{\dag}_{-\frac{1}{2}t}],\nonumber\\
H_{STP}(q^2)&=&Re[H_{\frac{1}{2}1}H^{\dag}_{-\frac{1}{2}t}-H_{-\frac{1}{2}-1}H^{\dag}_{\frac{1}{2}t}],\nonumber\\
H_{LTP}(q^2)&=&Re[H_{\frac{1}{2}1}H^{\dag}_{-\frac{1}{2}0}-H_{-\frac{1}{2}-1}H^{\dag}_{\frac{1}{2}0}].
\end{eqnarray}
We will use these helicity amplitudes to calculate the desired physical quantities in terms of hadronic form factors in the next subsection.

\subsection{Angular distribution and observables}
We now consider the three-body cascade decay $\Xi_b(p) \rightarrow \Xi_c(p') W^-(q)(\rightarrow \ell^-(p_1) \bar{\nu}_\ell(p_2))$ with $q=p_1+p_2$, $p_1^2=m_\ell^2$, and $p_2^2=0$. After summing over the helicities of all particles, the angular decay distribution  can be described in terms of the invariant variable $q^2$ and the polar angle $\theta$  with $\theta$ being the angle of the lepton in the $W$ rest frame with respect to the $W$ momentum. The twofold angular differential distribution reads~\cite{Gutsche:2015mxa, Faustov:2016pal, Bialas:1992ny}
\begin{eqnarray}
\frac{d^2\Gamma_\ell}{dq^2d\cos \theta}=\frac{G_F^2|V_{cb}|^2q^2|P|}{192M^2\pi^3}(1-\frac{m_\ell^2}{q^2})^2W(\theta),
\end{eqnarray}
where $|P|=\frac{\sqrt{(Q_+^2-q^2)(Q_-^2-q^2)}}{2M}$ is the momentum of the outgoing baryon and $m_\ell$ is the lepton mass with $\ell=e,\mu,\tau$. Moreover, we adopt $m_\tau=1.777$ GeV and neglect the masses of muons and electrons.  The $\theta$-dependent term $W(\theta)$ takes the form~\cite{Gutsche:2015mxa}
\begin{eqnarray}\label{eq:w}
W(\theta)&=&\frac{3}{8}[1+\cos^2(\theta)]H_T-\frac{3}{4}\cos(\theta)H_{TP}+\frac{3}{4}\sin^2(\theta)H_L \nonumber\\&&
+\frac{m_\ell^2}{2q^2}[\frac{3}{2}H_S+\frac{3}{4}\sin^2(\theta)H_T+\frac{3}{2}\cos^2(\theta)H_L-3\cos \theta H_{SL}],
\end{eqnarray}
where the structure functions $H_i$ have been specified in Eqs.~(\ref{eq:H1}) and~(\ref{eq:H2}). After integrating Eq.~(\ref{eq:w}) over the angular variable, one can obtain the total helicity amplitude
\begin{eqnarray}
H=\int d \cos \theta W(\theta)=H_T+H_L+\frac{m_\ell^2}{2q^2}(H_T+H_L+3H_S),
\end{eqnarray}
where the first two terms are non-spin-flip, while the last three terms  proportional to $m_\ell$ are lepton helicity flip  contributions. Then we can write the differential decay rate  in the following forms:
\begin{eqnarray}
\frac{d\Gamma_\ell}{dq^2}&=&\frac{G_F^2|V_{cb}|^2q^2|P|}{192M^2\pi^3}(1-\frac{m_\ell^2}{q^2})^2H,
\end{eqnarray}
and the branching ratio can be further obtained by integrating with respect to $q^2$
\begin{eqnarray}
\mathcal{B}_\ell&=&\tau_{\Xi_b}\int_{m_\ell^2}^{(M-m)^2} dq^2 \frac{d\Gamma_\ell}{dq^2},
\end{eqnarray}
with $\tau_{\Xi_b}$ denoting the lifetime of $\Xi_b$ baryon. Also, the differential and integrated ratios are defined as
\begin{eqnarray}
\mathcal{R}_{\Xi_c}(q^2)&=&\frac{d\Gamma_\tau/dq^2}{d\Gamma_e/dq^2},\\
\mathcal{R}_{\Xi_c}&=&\frac{\mathcal{B}_\tau}{\mathcal{B}_e}.
\end{eqnarray}
The forward-backward asymmetry of the charged lepton, which denotes the difference between the number of particles produced in the forward and backward directions, can be defined as
\begin{eqnarray}\label{eq:AFB}
A_{FB}(q^2)=\frac{\int_{0}^1 d \cos \theta W(\theta)-\int_{-1}^0d \cos \theta W(\theta) }
{\int_{0}^1d \cos \theta W(\theta)
+\int_{-1}^0 d \cos \theta W(\theta)}=-\frac{3}{4H}(H_{TP}+\frac{2m_\ell^2}{q^2}H_{SL}).
\end{eqnarray}
The term quadratic in $\cos \theta$ in the distribution Eq.~(\ref{eq:w}) is the convexity parameter defined by
\begin{eqnarray}
C_{F}(q^2)= \frac{1}{H}\frac{d^2W(\theta)}{d\cos^2\theta}=\frac{3}{4}(1-\frac{m_\ell^2}{q^2})\frac{H_T-2H_L}{H},
\end{eqnarray}
which gives the angular decay distribution over $\cos \theta$ dependence. The polarization components of the final baryon $\Xi_c$ can be obtained from the spin density matrix of the daughter baryon. One has~\cite{Gutsche:2015mxa}
\begin{eqnarray}
P^{h}_z(q^2)&=& \frac{H_{TP}+H_{LP}+\frac{m_\ell^2}{2q^2}(H_{TP}+H_{LP}+3H_{SP})}{H}, \\
P^{h}_x(q^2)&=& -\frac{3\pi}{4\sqrt{2}} \frac{H_{LT}-\frac{m_\ell^2}{q^2}H_{STP}}{H}.
\end{eqnarray}
The longitudinal and transverse polarization components of the charged lepton read,
\begin{eqnarray}\label{eq:pl}
P^{\ell}_z(q^2)&=& -\frac{H_{T}+H_{L}-\frac{m_\ell^2}{2q^2}(H_{T}+H_{L}+3H_{S})}{H}, \\
P^{\ell}_x(q^2)&=& -\frac{3\pi}{4\sqrt{2}}\sqrt{\frac{m_\ell^2}{2q^2}}\frac{H_{TP}-2H_{SL}}{H},
\end{eqnarray}
respectively. Note that when calculating the $q^2$ averages, one has to remember to include the $q^2$-dependent factor  $q^2|P|(1-\frac{m_\ell^2}{q^2})^2$ in the numerator and denominator of the relevant asymmetry expressions. Both the numerator and denominator should be integrated separately over the permissible phase space, $q^2\in [m_\ell^2,(M-m)^2]$.

\section{Numerical results}\label{sec:results}
We first present all the inputs that are relevant for our numerical computation. The default values are quoted from the Particle Data Group (PDG)~\cite{pdg2024}
 \begin{eqnarray}
 M_{\Xi^-_b}&=&5.797~\text{GeV}, \quad m_{\Xi_c}=2.47~\text{GeV},\quad m_\tau=1.777~\text{GeV}, \quad\tau_{\Xi^-_b}=1.572~\text{ps},\nonumber\\
\lambda &=&0.22650, \quad  A=0.790,  \quad \bar{\rho}=0.141, \quad \bar{\eta}=0.357.
\end{eqnarray}
\subsection{ Form factors}
As is known that the PQCD calculations for the form factors are reliable in the low $q^2$ range. We need to extrapolate these form factors into the whole physically accessible region for studying their $q^2$ dependencies~\footnote{It was pointed out in Ref~\cite{Albertus:2005ud} that the Omn\`{e}s dispersion relation of the form factors can be used to combine predictions from various methods in different $q^2$ regions, which considerably diminishes the form-factor dependence on the nonperturbative strong-interaction effects at high energies. It would be advantageous to combine the reliable result for the high $q^2$ regions from the lattice computations using the Omn\`{e}s dispersion relation to improve the accuracy of the form factors in the future.}.
For this purpose, we use the $z$-series parametrization that is proposed in~\cite{Li:2022hcn},  
\begin{eqnarray}
z(q^2)=\frac{\sqrt{Q^2_+-q^2}-\sqrt{Q^2_+-Q^2_0}}{\sqrt{Q^2_+-q^2}+\sqrt{Q^2_+-Q^2_0}},
\end{eqnarray}
with $Q^2_0=Q^2_+(1-\sqrt{1-Q^2_-/Q^2_+})$. Keeping the series expansion of the form factors to the first power of the $z$-parameter, the parametrization of the form factors in the  whole physical region has the following form
\begin{eqnarray}\label{eq:Fq}
F(q^2)= \frac{a_0+a_1z(q^2)}{1-q^2/M^2_{\text{pole}}},
\end{eqnarray}
where the pole masses are given as~\cite{Faustov:2018ahb},
\begin{eqnarray}
M_{\text{pole}}= \left\{
            \begin{array}{ll}
               6.333~\text{GeV}, & \text{for}~ f_{1,2} \\
               6.743~\text{GeV}, & \text{for}~ g_{1,2} \\
               6.699~\text{GeV}, & \text{for}~ f_{3} \\
               6.275~\text{GeV}, & \text{for}~ g_{3} ,\\
            \end{array}
          \right.
\end{eqnarray}
and $a_{0,1}$ are free parameters  needed to be fitted.

\begin{table}[!htbh]
	\caption{The fitted parameters for the $\Xi_b\rightarrow \Xi_c$ transition form factors with different models of baryonic LCDAs.}
	\label{tab:form}
	\begin{tabular}[t]{lccccccc}
	\hline\hline
Model   & Parameter     &$f_1$ &$f_2$ &$f_3$ &$g_1$ &$g_2$ &$g_3$\\ \hline
Exponential &$a_0 $     &0.611  &0.018  &0.017  &0.624   &-0.017  &-0.017\\
            &$a_1 $     &-3.159 &-0.312 &-0.266 &-3.873  &0.265   &0.289\\
QCDSR       &$a_0 $     &0.623  &0.019  &0.018  &0.635   &-0.018  &-0.018\\
            &$a_1 $     &-3.051 &-0.345 &-0.277 &-3.641  &0.276   &0.316\\
Gegenbauer&$a_0 $     &0.671  &0.042  &0.007  &0.671   &-0.007  &-0.040\\
            &$a_1 $     &-5.152 &-0.816 &-0.074 &-5.050  &0.052   &0.790\\
		\hline\hline
	\end{tabular}
\end{table}

\begin{table}[!htbh]
\caption{ Comparisons of the form factors $f_i(q^2)$   and $g_i(q^2)$   at $q^2=0$ and $q^2=q^2_{\text{max}}$ from three different LCDAs models obtained in
this work and previous studies in different literature.  }
	\label{tab:form1}
	\begin{tabular}[t]{lcccc}
	\hline\hline
      Model                &  Form factor     &$i=1$ &$i=2$ &$i=3$\\ \hline
This work (Exponential model)     &$f_i(q^2=0)$ &$0.545^{+0.292+0.066+0.012}_{-0.160-0.046-0.037}$ &$0.011^{+0.005+0.001+0.001}_{-0.003-0.000-0.002}$ &$0.011^{+0.004+0.001+0.001}_{-0.002-0.001-0.001}$\\
                                  &$g_i(q^2=0)$ &$0.542^{+0.310+0.073+0.011}_{-0.162-0.040-0.031}$ &$-0.011^{+0.003+0.001+0.001}_{-0.004-0.001-0.001}$ &$-0.010^{+0.002+0.000+0.001}_{-0.005-0.002-0.002}$\\
                                  &$f_i(q^2=q^2_{\text{max}})$ &$0.940^{+0.501+0.124+0.007}_{-0.283-0.049-0.042}$ &$0.034^{+0.017+0.003+0.004}_{-0.009-0.001-0.002}$  &$0.030^{+0.011+0.004+0.001}_{-0.005-0.002-0.002}$\\
                                  &$g_i(q^2=q^2_{\text{max}})$&$0.938^{+0.414+0.075+0.005}_{-0.279-0.084-0.063}$ &$-0.030^{+0.006+0.002+0.003}_{-0.010-0.003-0.003}$ &$-0.032^{+0.007+0.002+0.002}_{-0.015-0.002-0.004}$\\
This work (QCDSR model)            &$f_i(q^2=0)$     &$0.554^{+0.038+0.040+0.006}_{-0.012-0.010-0.037}$ &$0.012^{+0.000+0.001+0.001}_{-0.001-0.001-0.001}$  &$0.012^{+0.001+0.000+0.002}_{-0.000-0.000-0.001}$\\
                                   &$g_i(q^2=0)$   &$0.552^{+0.041+0.035+0.019}_{-0.012-0.018-0.034}$ &$-0.012^{+0.000+0.001+0.001}_{-0.001-0.001-0.001}$ &$-0.011^{+0.001+0.000+0.001}_{-0.001-0.001-0.001}$\\
                                   &$f_i(q^2=q^2_{\text{max}})$&$0.954^{+0.066+0.033+0.067}_{-0.024-0.034-0.037}$ &$0.037^{+0.0018+0.000+0.003}_{-0.001-0.001-0.003}$  &$0.032^{+0.003+0.002+0.004}_{-0.000-0.000-0.002}$\\
                                   &$g_i(q^2=q^2_{\text{max}})$   &$0.946^{+0.054+0.041+0.022}_{-0.020-0.012-0.041}$ &$-0.032^{+0.001+0.000+0.003}_{-0.001-0.000-0.004}$ &$-0.034^{+0.001+0.000+0.003}_{-0.001-0.000-0.003}$\\
This work (Gegenbauer model)    &$f_i(q^2=0)$      &$0.560^{+0.039+0.041+0.028}_{-0.000-0.000-0.038}$ &$0.024^{+0.001+0.001+0.003}_{-0.001-0.000-0.002}$  &$0.006^{+0.002+0.002+0.000}_{-0.002-0.001-0.001}$\\
                                  &$g_i(q^2=0)$    &$0.567^{+0.034+0.029+0.010}_{-0.000-0.006-0.039}$ &$-0.005^{+0.002+0.001+0.000}_{-0.002-0.002-0.001}$ &$-0.023^{+0.000+0.000+0.001}_{-0.002-0.001-0.003}$\\
                                  &$f_i(q^2=q^2_{\text{max}})$ &$1.084^{+0.109+0.088+0.071}_{-0.000-0.024-0.073}$ &$0.082^{+0.008+0.007+0.011}_{-0.002-0.001-0.007}$  &$0.012^{+0.005+0.006+0.001}_{-0.007-0.007-0.000}$\\
                                  &$g_i(q^2=q^2_{\text{max}})$    &$1.034^{+0.067+0.107+0.079}_{-0.000-0.000-0.060}$ &$-0.010^{+0.005+0.005+0.001}_{-0.005-0.006-0.001}$ &$-0.080^{+0.002+0.000+0.008}_{-0.006-0.005-0.010}$\\
  NRQM~\cite{ Cheng:1996cs}       &$f_i(q^2=0)$      &0.533 &0.124 &-0.018\\
                                  &$g_i(q^2=0)$     &0.580 &0.019 &-0.135\\
LFQM~\cite{Li:2021kfb}             &$f_i(q^2=0)$      &0.481 &0.127 &-0.046\\
                                  &$g_i(q^2=0)$     &0.471 &0.026 &-0.154\\
                                   &$f_i(q^2=q^2_{\text{max}})$       &1.015 &0.312 &-0.097\\
                                   &$g_i(q^2=q^2_{\text{max}})$     &0.978 &0.068 &-0.377\\
LFQM~\cite{Ke:2024aux}            &$f_i(q^2=0)$       &0.467 &0.185 & $\cdots$\\
                                  &$g_i(q^2=0)$     &0.448 &0.052 &$\cdots$\\
LF~\cite{Zhao:2018zcb}           &$f_i(q^2=0)$        &0.654 &0.143 & $\cdots$ \\
                                  &$g_i(q^2=0)$     &0.640 &0.015 &$\cdots$\\
LFQM~\cite{Chua:2019yqh}          &$f_i(q^2=0)$       &0.437 &0.123 &0.057\\
                                  &$g_i(q^2=0)$     &0.429 &-0.059 &-0.145\\
                                  &$f_i(q^2=q^2_{\text{max}})$ &0.714 &0.206 &0.102 \\
                                  &$g_i(q^2=q^2_{\text{max}})$      &0.693 &-0.099 &-0.214\\
RQM~\cite{Faustov:2018ahb}       &$f_i(q^2=0)$        &0.474 &0.150 & $0.081$ \\
                                  &$g_i(q^2=0)$     &0.449 &-0.030 &-0.285\\
                                  &$f_i(q^2=q^2_{\text{max}})$&0.945 &0.426 & $0.161$ \\
                                  &$g_i(q^2=q^2_{\text{max}})$      &0.962 &-0.104 &-0.752\\
QCDSR~\cite{Zhao:2020mod}      &$f_i(q^2=0)$          &0.500 &0.141 & $0.016$ \\
                                   &$g_i(q^2=0)$    &0.500 &-0.035 &-0.191\\
                              &$f_i(q^2=q^2_{\text{max}})$ &0.971 &0.346 & $0.070$ \\
                             &$g_i(q^2=q^2_{\text{max}})$&1.107 &-0.125 &-0.519\\
HQET~\cite{Manohar:2000dt}  &$f_i(q^2=q^2_{\text{max}})$&1.029 &0.271 & $-0.050$ \\
                           &$g_i(q^2=q^2_{\text{max}})$ &1.029 &0.050 &-0.271\\
		\hline\hline
	\end{tabular}
\end{table}

\begin{figure}[!htbh]
\begin{center}
\setlength{\abovecaptionskip}{0pt}
\centerline{
\hspace{-1.0cm}\subfigure{ \epsfxsize=7cm \epsffile{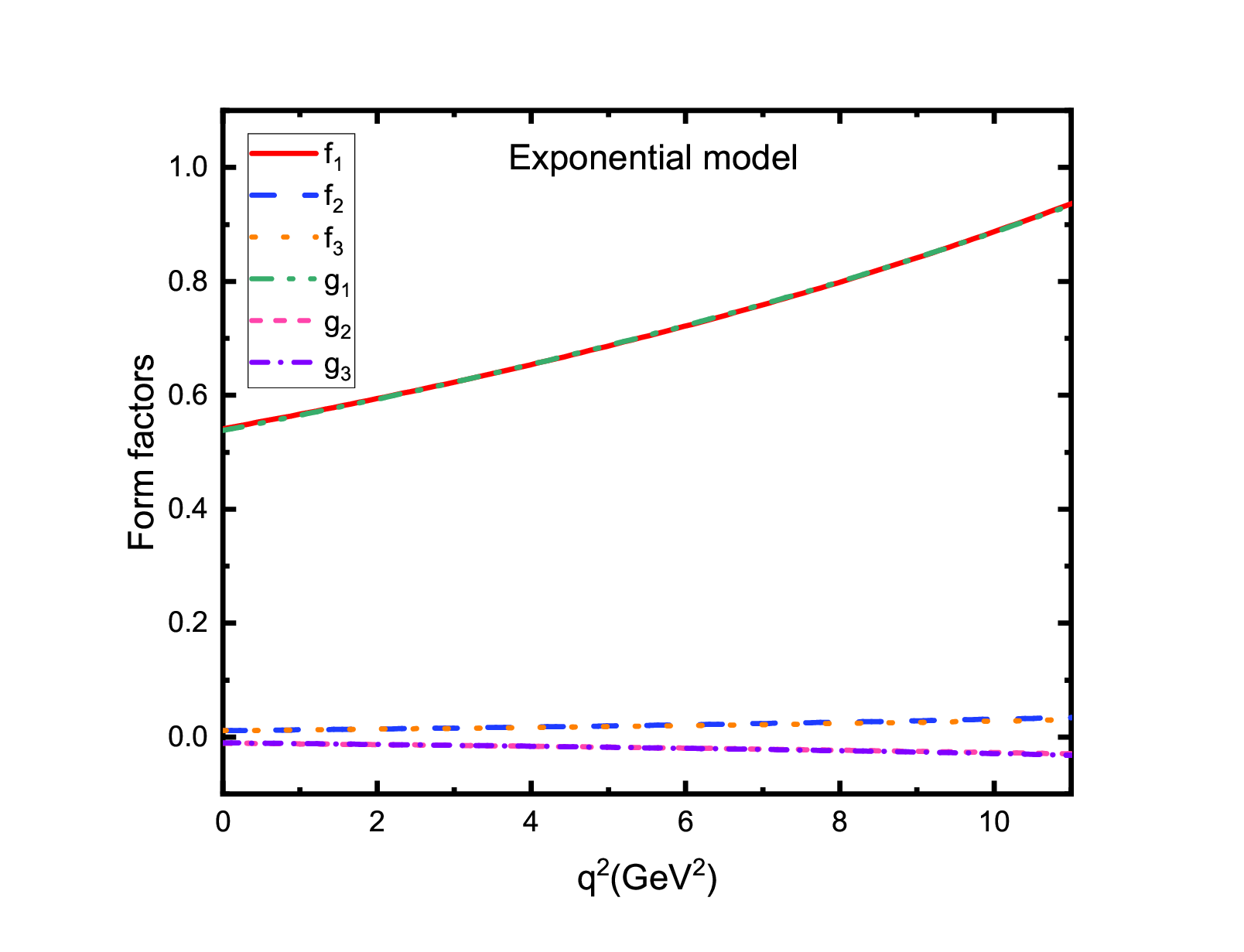}}
\hspace{-1.0cm}\subfigure{ \epsfxsize=7cm \epsffile{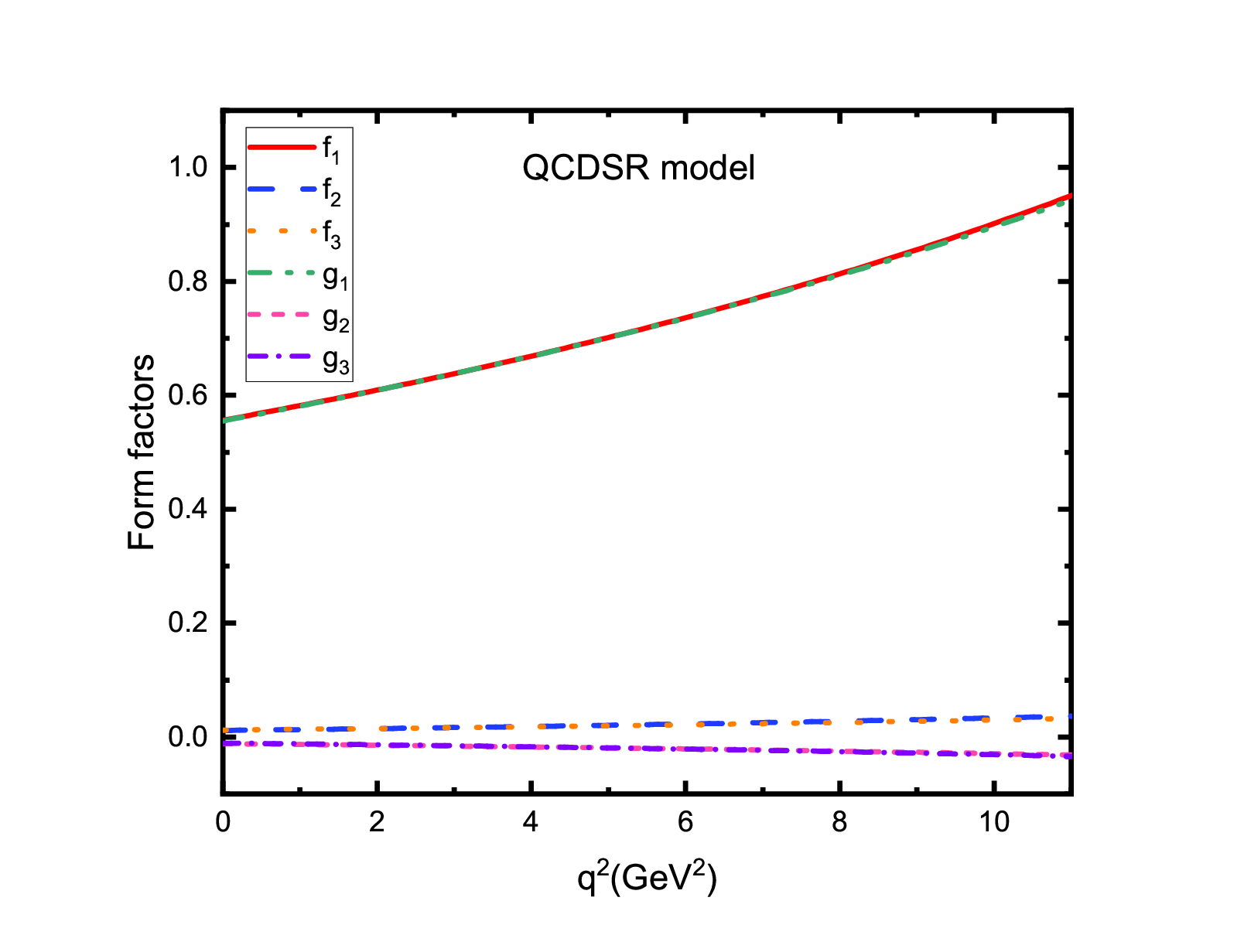}}
\hspace{-1.0cm}\subfigure{ \epsfxsize=7cm \epsffile{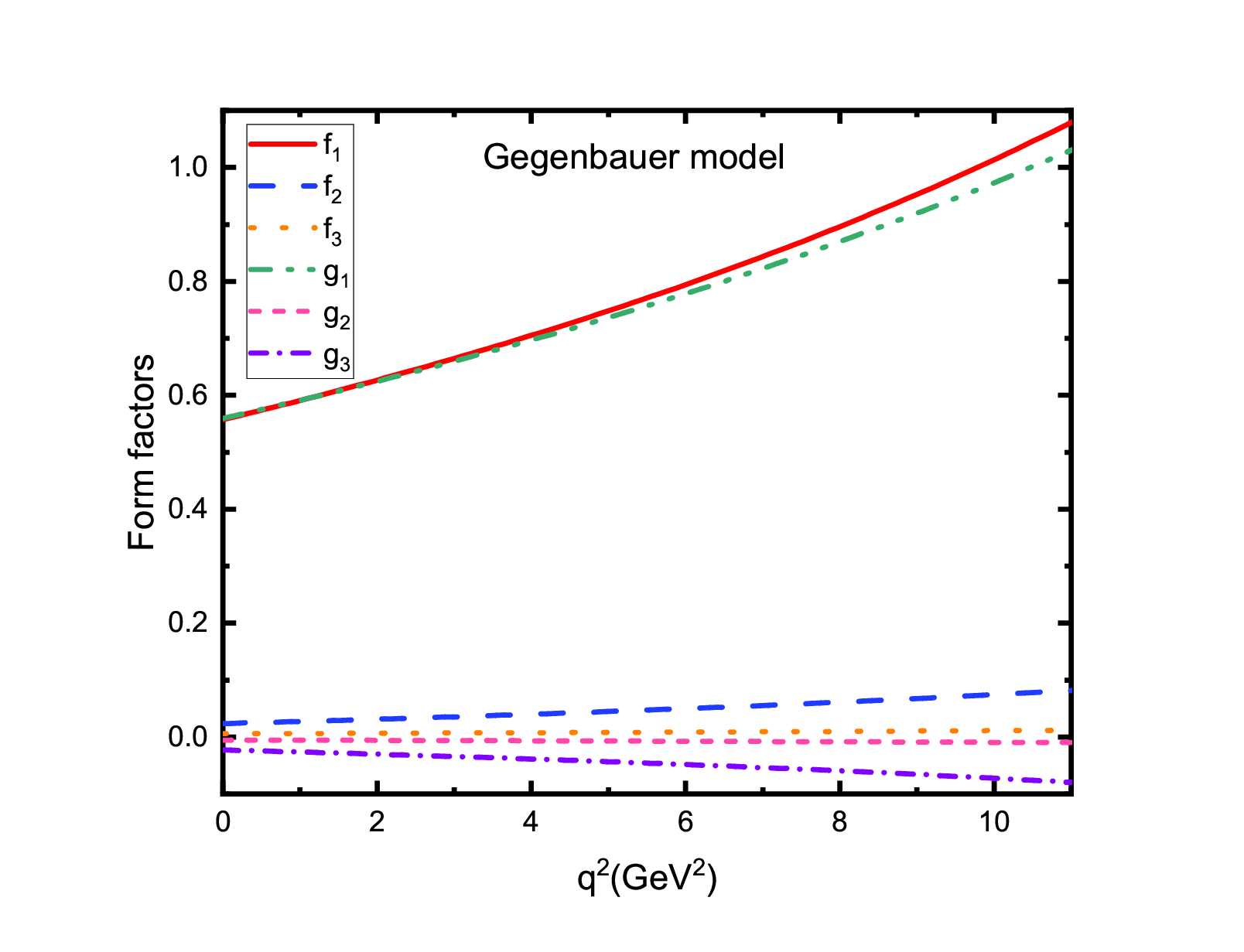}}}
\vspace{1cm}
\caption{$q^2$ dependencies of the form factors with three different models of baryonic LCDAs.}
 \label{fig:Form}
\end{center}
\end{figure}

To access the $q^2$ dependence of the form factors, we first numerically evaluate each  form factor  in Eq.~(\ref{eq:FG}) at ten $q^2$ values from 0 to $m^2_\tau$ region. Then we performed a ten-point fit using Eq.~(\ref{eq:Fq}) to determine the parameters $a_0$ and $a_1$. The fitted central values of $a_0$ and $a_1$ with three different LCDAs models are given separately in Table~\ref{tab:form}. The $q^2$  behaviors of the form factors in the allowed region are illustrated in Fig.~\ref{fig:Form}. As it is expected from weak decays, the form factors demonstrate a good behavior that their magnitudes grow gradually with increasing $q^2$. $f_1(q^2)$ and  $g_1(q^2)$ show similar $q^2$  dependence  and dominate over other four form factors, while $f_{2,3}(q^2)$ and $g_{2,3}(q^2)$  exhibit relatively low sensitivity to variations in $q^2$. The results of $f_1(q^2)$ and $g_1(q^2)$ from all three models of LCDAs are highly consistent. We also observe that $f_{2,3}(q^2)$ and $g_{2,3}(q^2)$ from the Exponential model and QCDSR model of baryonic LCDAs are of similar sizes but opposite in signs across the entire $q^2$ range. $f_3(q^2)$ and $g_2(q^2)$ from the Gegenbauer model are nearly flat and smallest, which are expected to be vanishing in the exact SU(3) symmetry limit~\cite{Sasaki:2008ha}. In general, the predicted form factors follow the relations postulated in HQET $f_1(q^2)\approx g_1(q^2)$ and $f_{2,3}(q^2) \approx g_{2,3}(q^2) \approx 0$~\cite{Mannel:1990vg}. The fit functions of form factors will be employed as the primary input parameters to evaluate the physical observables at different lepton channels in the subsequent subsection.

In Table~\ref{tab:form1}, we compare our results of the form factors at the $q^2=0$  and  $q^2=q^2_{\text{max}}$ end points of the $\Xi_b\rightarrow \Xi_c$ transition  with the other theoretical predictions~\cite{Cheng:1996cs, Li:2021kfb, Ke:2024aux, Zhao:2018zcb, Chua:2019yqh, Faustov:2018ahb, Zhao:2020mod,Manohar:2000dt}. Note that the definitions of the form factors in the literature are different. To compare these results directly,  we have rescaled them according to the definitions in Eq.~(\ref{eq:FFs}). The first theoretical error of our results comes from the variations of the relevant parameters in the $\Xi_b$ LCDAs. The second error arises from the same parameters, but for $\Xi_c$ LCDAs  and $\beta$ parameter have been added in quadrature. The last one corresponds to the hard scale $t$ varying from 0.8$t$ to 1.2$t$. It can be seen apparently that the predicted form factors with the Exponential model show greater sensitivity to variations due to uncertainties corresponding to the model parameters. The uncertainties in the parameters from the QCDSR model and Gegenbauer model are relatively smaller, leading to the changes in form factors being roughly less than $10\%$. In general,  the main errors belong to the uncertainties of the parameters in the LCDAs. Our predictions  are more sensitive to the $b$-baryon LCDAs than the charmed ones. From Table~\ref{tab:form1}, we see most of the theoretical calculations predict comparable values for the form factors $f_1(0)$ and $g_1(0)$ ranging from 0.43 to 0.65. Our results are more close to  the predictions from QCD  sum rule (QCDSR)~\cite{Zhao:2020mod} and the nonrelativistic quark model (NRQM)~\cite{Cheng:1996cs}, but slightly larger than the values from the light-front quark model (LFQM)~\cite{Li:2021kfb,Ke:2024aux,Chua:2019yqh}, and relativistic quark-diquark model (RQM)~\cite{Faustov:2018ahb}. The results from  the light-front approach (LF)~\cite{Zhao:2018zcb} are somewhat large. All of these approaches predict  the small values of $f_3(0)$ and $g_2(0)$ but there exists an uncertainty about their sign. We obtain positive $f_{2,3}(q^2)$ and negative $g_{2,3}(q^2)$ values that are the same as  those obtained by LFQM~\cite{Chua:2019yqh}, the RQM~\cite{Faustov:2018ahb}, and the QCDSR~\cite{Zhao:2020mod}, but different from NRQM~\cite{ Cheng:1996cs} and those derived by the LFQM method in Refs.~\cite{Li:2021kfb,Ke:2024aux,Zhao:2018zcb}. A similar situation also exists in the corresponding values at zero recoil. The PQCD predictions on the magnitudes of $f_{2,3}(q^2)$ and $g_{2,3}(q^2)$  are generally within few $\%$, which are much smaller than other predictions but closer to the expectations in the heavy quark limit. In addition, most of the theoretical calculations on the $f_1(q^2_{\text{max}})$ and  $g_1(q^2_{\text{max}})$ are in nice agreement with the expectation in heavy quark limit with $f_1(q^2_{\text{max}})=g_1(q^2_{\text{max}})=1$~\cite{Chua:2019yqh}. Nevertheless, the corresponding numbers from LFQM~\cite{Chua:2019yqh}  are somewhat lower than the expectation by roughly $30\%$. We expect more theoretical works on these form factors to further enrich our knowledge on these weak decays.

\subsection{\texorpdfstring{$q^2$}{}-dependent observables}
\begin{figure}[!htbh]
\begin{center}
\centerline{
\hspace{0cm}\subfigure{\epsfxsize=8 cm \epsffile{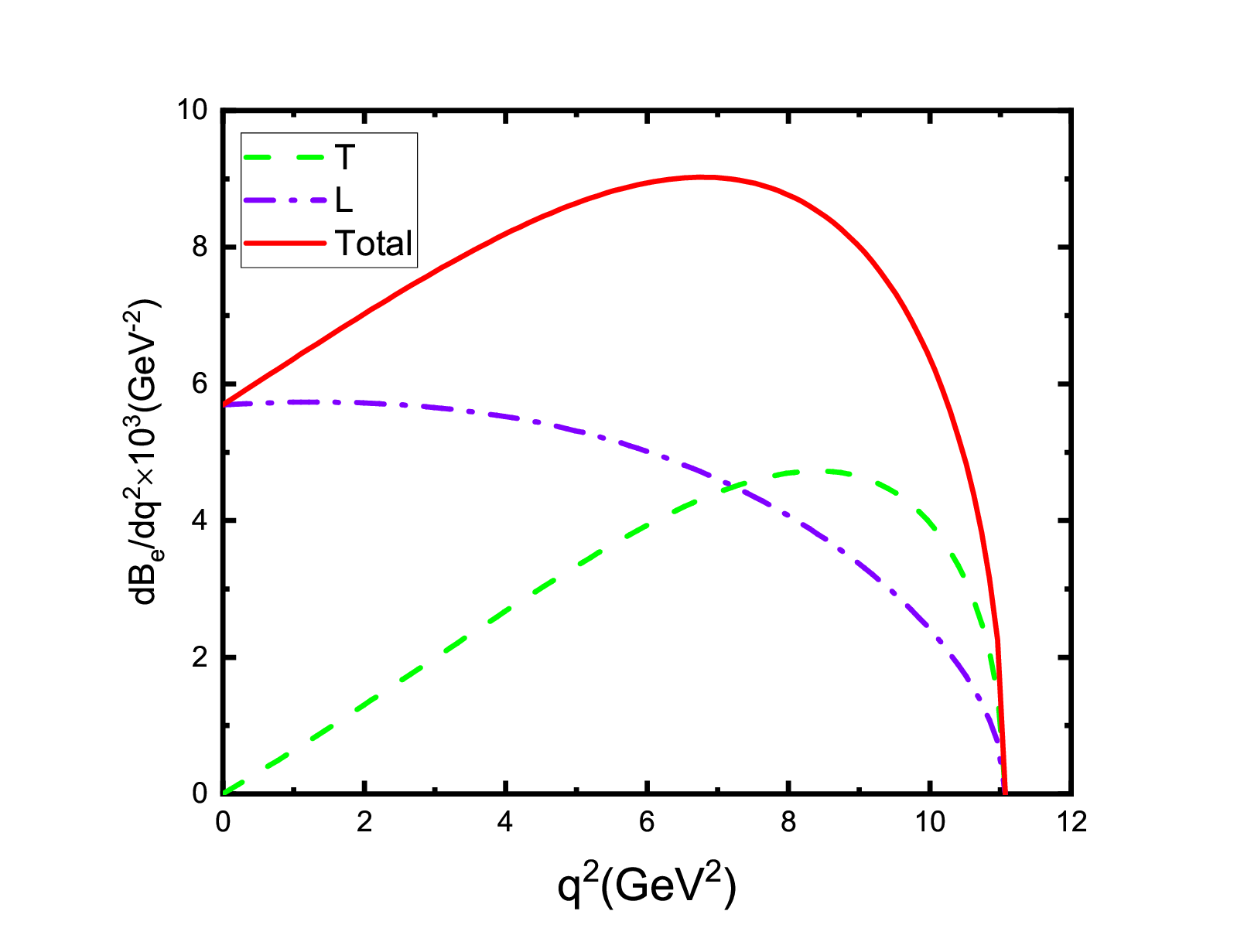} }
\hspace{0cm}\subfigure{ \epsfxsize=8 cm \epsffile{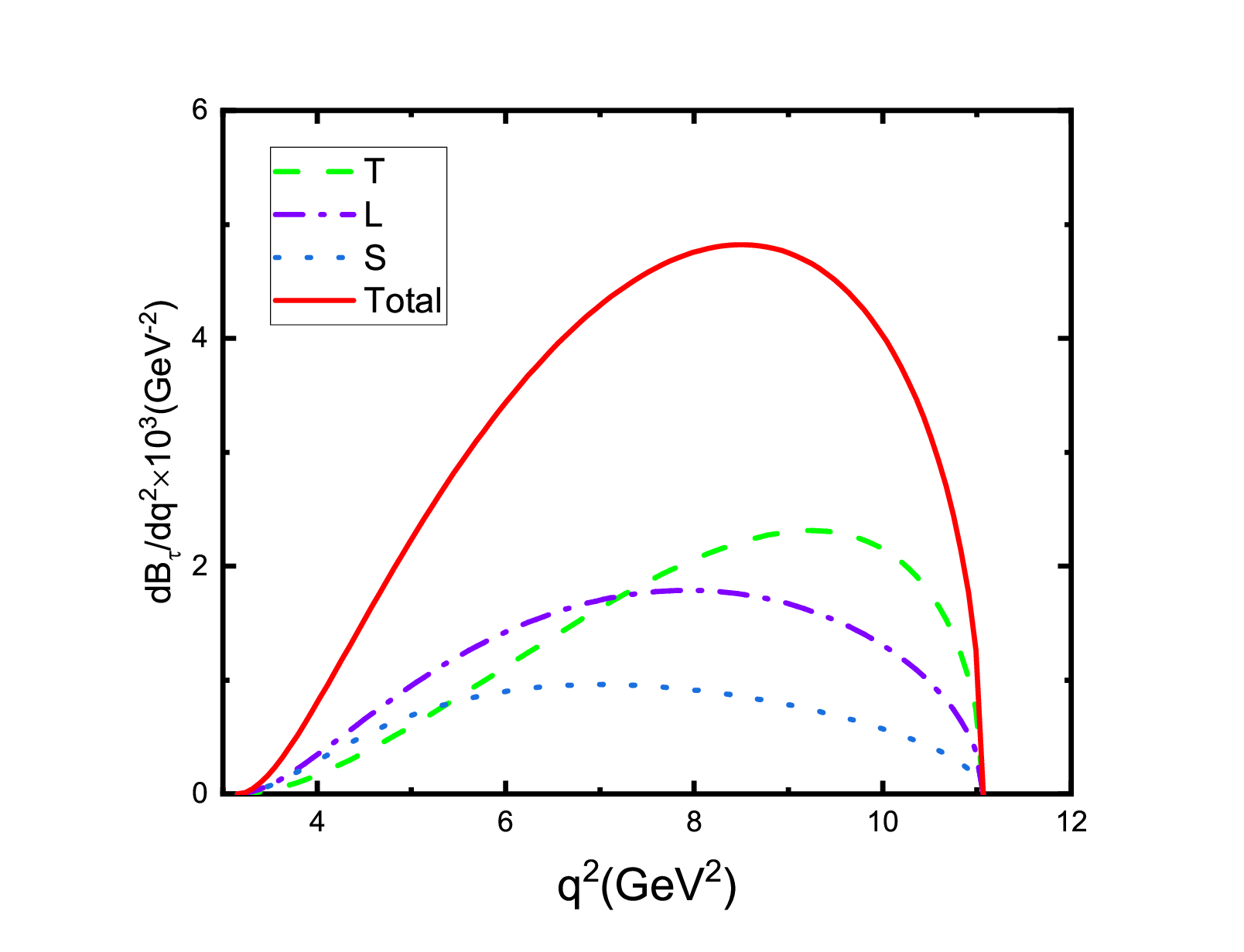}}}
\vspace{0cm}\caption{ The differential decay branching ratios of the semileptonic $\Xi_b\rightarrow \Xi_c e^-\bar{\nu}_e$ (left)
and $\Xi_b\rightarrow \Xi_c \tau^-\bar{\nu}_\tau$ (right) decays  in the full $q^2$ kinematic  region.
The longitudinal, transverse, and scalar partial decay distributions are shown in dot dashed, dashed, and dot, respectively.
The solid curves are their sum.}
 \label{fig:dB}
\end{center}
\end{figure}

Taking the obtained form factors as input, we now calculate several $q^2$-dependent observables defined in the previous section for the semileptonic decays under consideration. Because the results based on the three models of baryonic LCDAs  are similar, we will take the Gegenbauer model as an example to discuss the $q^2$ dependencies of physical quantities. The $q^2$  variation of differential branching ratios ($d\mathcal{B}(q^2)/dq^2$), the leptonic forward-backward asymmetries ($A_{FB}(q^2)$), the convexity parameter ($C_F(q^2)$), the final hadron polarizations ($P^h(q^2)$), and the lepton polarizations ($P^l(q^2)$) for the $\tau$ and $e$ modes are plotted in Figs.~\ref{fig:dB}-\ref{fig:Plzx}, respectively. Since the electron and muon are very light compared with the heavy tau lepton, we neglect their masses in the calculations. In this case, the results for the muon mode are almost identical to those of the electron mode and will not be displayed separately. We list our major findings as follows:
\begin{itemize}
\item
In Fig.~\ref{fig:dB}, we display  the $q^2$ distributions of longitudinal, transverse, and scalar differential branching ratios as well as their total contribution. It is easy to see that each of the curves  increases initially with the increasing of the $q^2$ and after reaching a maximum it starts to decrease. For both the  $e$ and $\tau$ channels, we observe that the longitudinal component dominates in the low $q^2$ region, while the transverse one dominates in the high $q^2$ region. At the point $q^2=7.23$ GeV$^2$, the longitudinal and transverse polarizations are equal. The scalar helicity component is suppressed by the lepton mass with respect to the other two components; its contribution vanishes for the $e$ mode  but contributes significantly to the total rate in the $\tau$ mode. Compared to the $e$ mode, the differential branching ratios involved $\tau$ have lower peaks due to the smaller phase space available.

\begin{figure}[!htbh]
\begin{center}
\centerline{
\hspace{0cm}\subfigure{\epsfxsize=8cm \epsffile{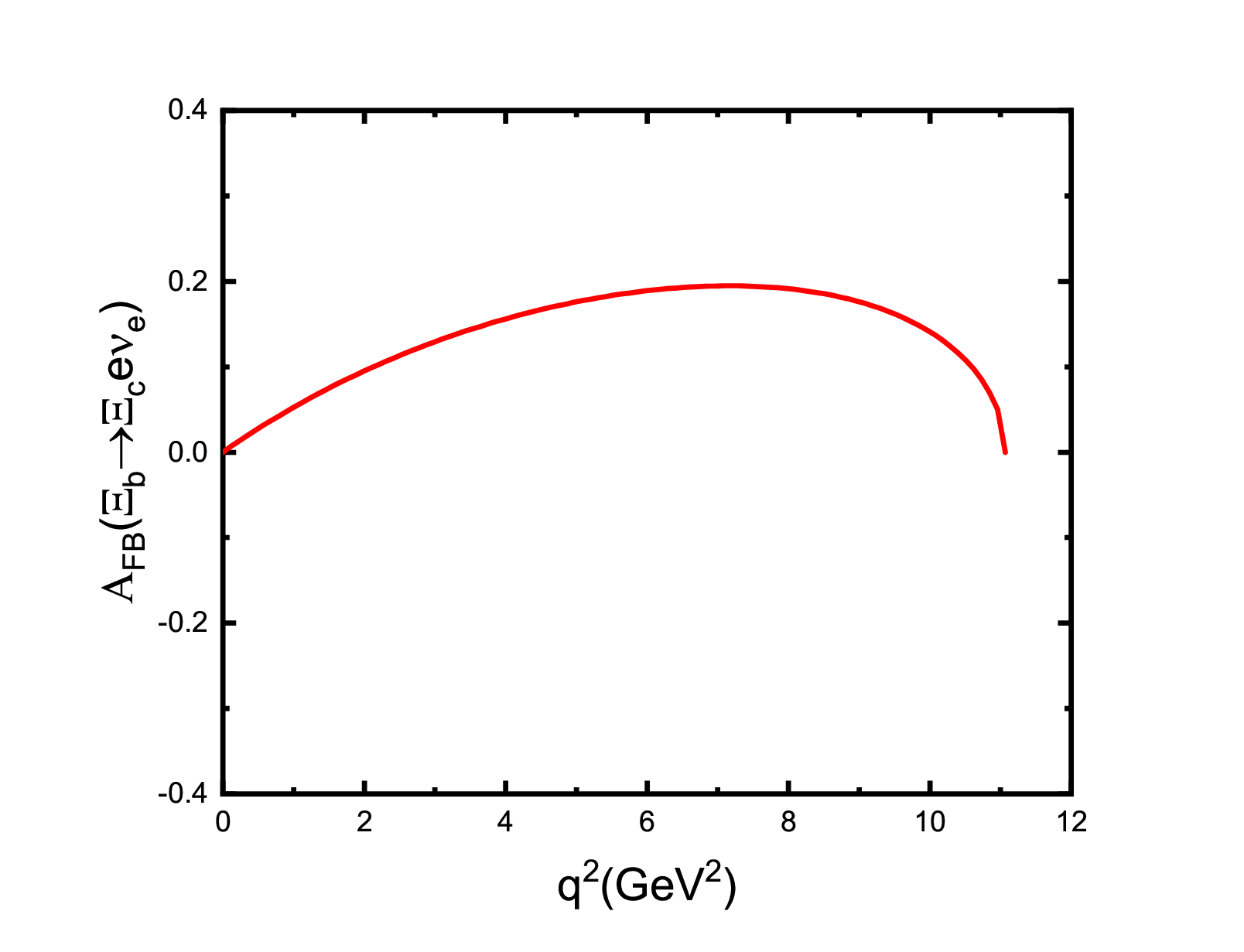} }
\hspace{0cm}\subfigure{ \epsfxsize=8 cm \epsffile{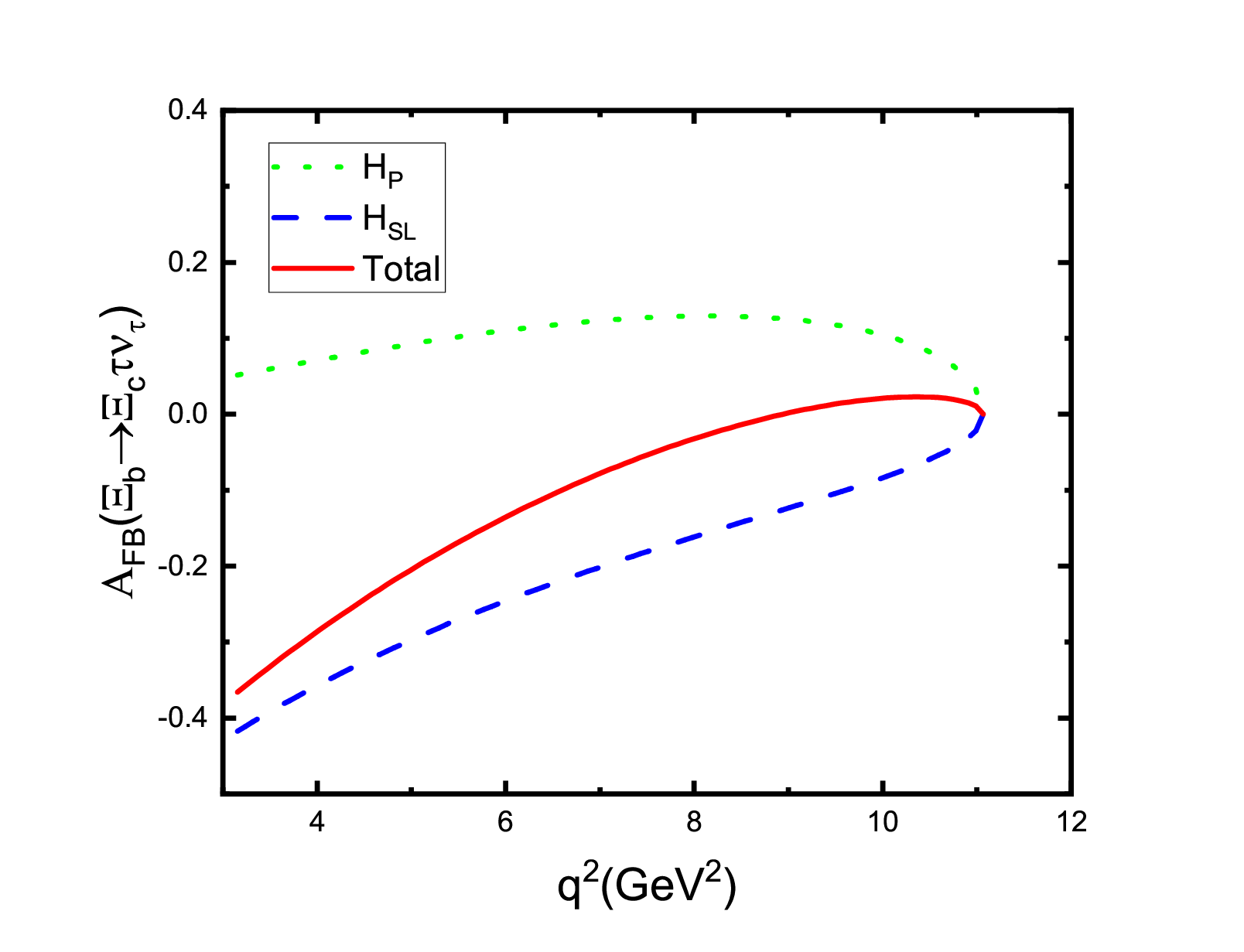}}}
\vspace{0cm}\caption{
  The dependence of the leptonic forward-backward asymmetries on $q^2$ for the  $\Xi_b\rightarrow \Xi_c e^-\bar{\nu}_e$ (left) and $\Xi_b\rightarrow \Xi_c \tau^-\bar{\nu}_\tau$ (right) transitions.}
 \label{fig:AFB}
\end{center}
\end{figure}

\item
From Fig.~\ref{fig:AFB}, we observe that the forward-backward asymmetry $A_{\text{FB}}$ for the $e$ mode approaches to zero both at zero recoil and maximum recoil. Beyond that, it is positive for other kinematic ranges, which reflect the dominance of the parity-violating helicity structure function, $H_{TP}$, with a substantial contribution from $H_{-\frac{1}{2}-1}$ amplitude. However, the taunic mode receives an additional contribution from the scalar-longitudinal interference ($H_{SL}$), which affects the asymmetry significantly in the small $q^2$ region as shown in Fig.~\ref{fig:AFB}. This leads to $A^{\tau}_{\text{FB}}$ has a negative value at the low $q^2$ region and becomes positive at large recoil, and zero crossing occurs at $q^2=8.9$ GeV$^2$. We also observe the asymmetry takes the value $A^{\tau}_{\text{FB}}=-0.366$ at $q^2=m_{\tau}^2$ and $A^{\tau}_{\text{FB}}=0$ at $q^2=q^2_{\text{max}}$. The obtained behaviors of the forward-backward asymmetries for both $e$ and $\tau$ modes are quite close to the results in~\cite{Faustov:2018ahb}.

\begin{figure}[!htbh]
\begin{center}
\centerline{
\hspace{0cm}\subfigure{\epsfxsize=8cm \epsffile{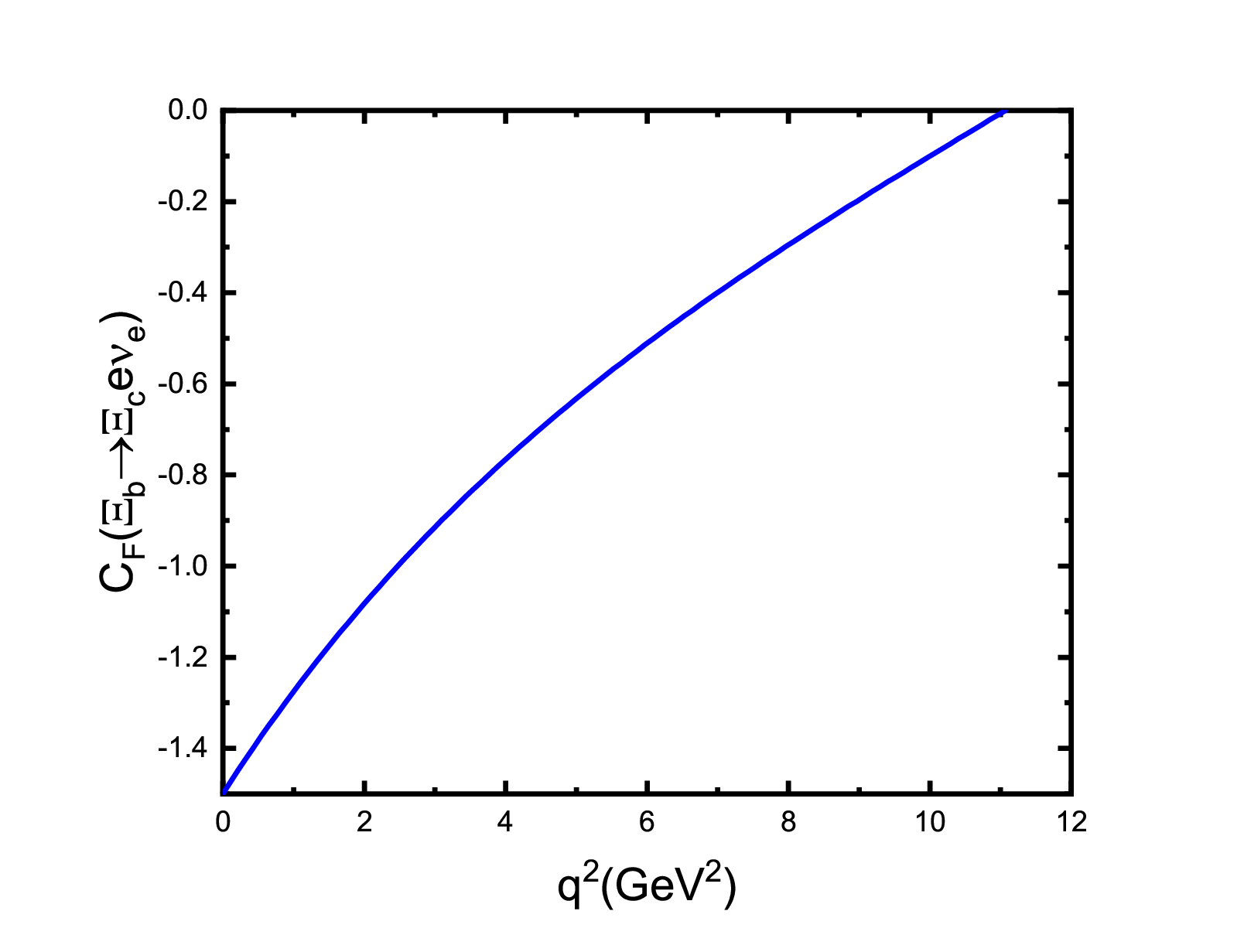} }
\hspace{0cm}\subfigure{ \epsfxsize=8 cm \epsffile{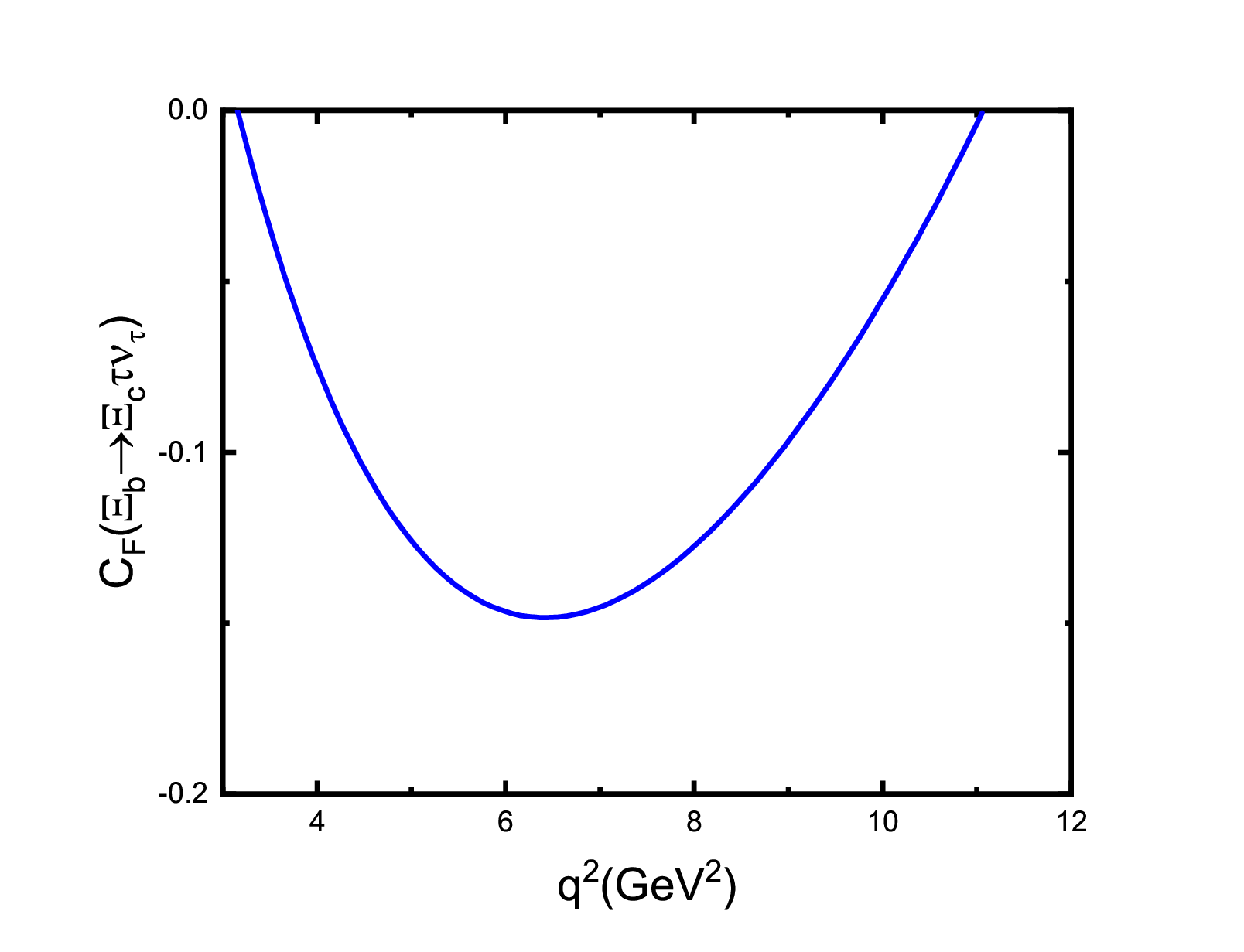}}}
\vspace{0cm}\caption{
  The dependence of the convexity parameter on $q^2$ for the  $\Xi_b\rightarrow \Xi_c e^-\bar{\nu}_e$ (left) and $\Xi_b\rightarrow \Xi_c \tau^-\bar{\nu}_\tau$ (right) transitions.}
 \label{fig:CF}
\end{center}
\end{figure}

\item
It is interesting to note that  $C_F(q^2)$ stays negative for both the $e$ and $\tau$ channels  throughout the available $q^2$ indicating a predominance of $H_L$. Again, the convexity parameter  has zero value for both the decay modes at zero recoil point. However, at maximum recoil, $C_F(q^2)$ becomes a large and negative value, -1.5, for the $e$ mode. We also observe the convexity parameter for the tauonic mode  remains very small in magnitude throughout the whole $q^2$ region and reaches its minimum value, -0.14 at $q^2=6.35$ GeV$^2$.

\begin{figure}[!htbh]
\begin{center}
\centerline{
\hspace{0cm}\subfigure{\epsfxsize=8cm \epsffile{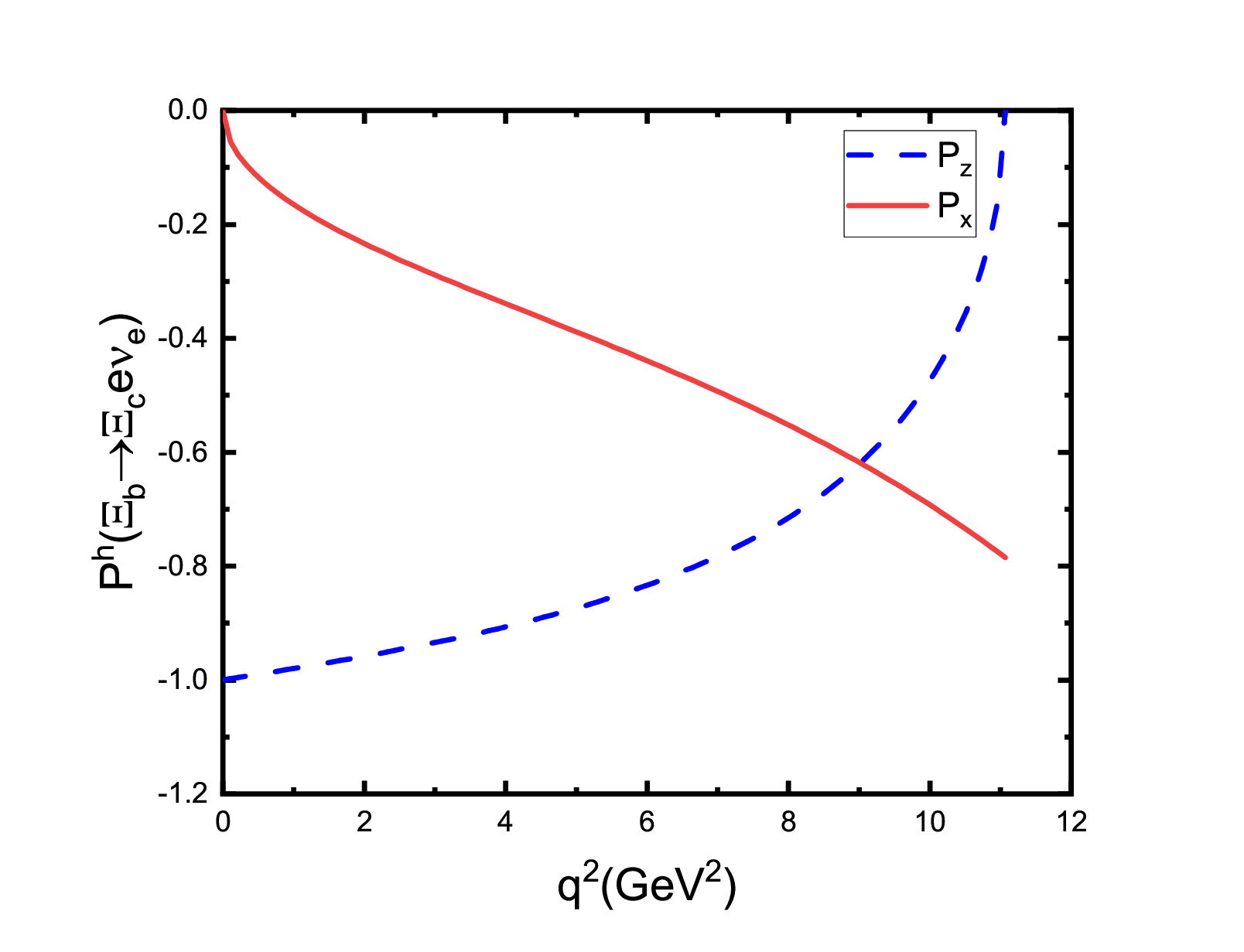} }
\hspace{0cm}\subfigure{ \epsfxsize=8 cm \epsffile{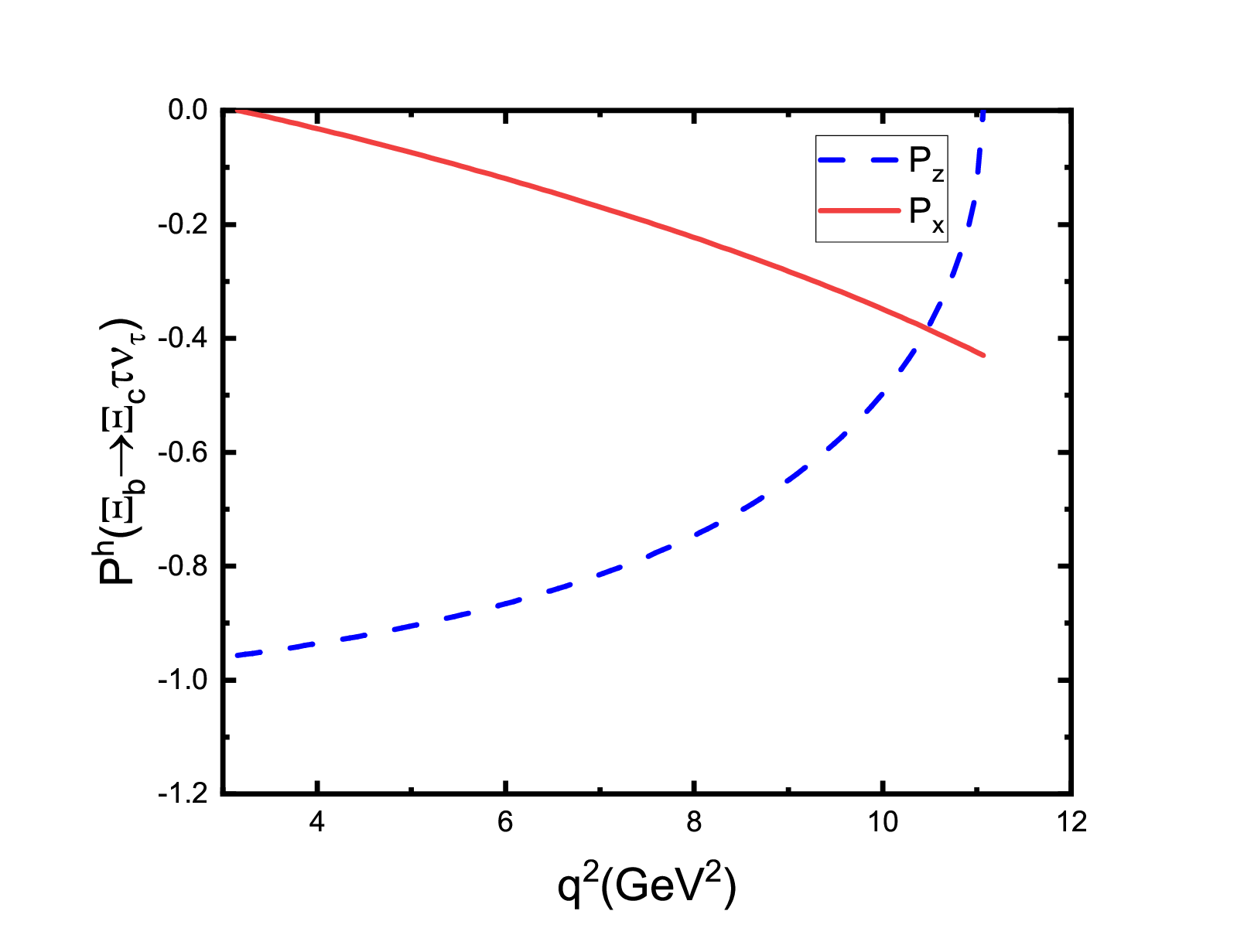}}}
\vspace{0cm}\caption{
  The dependence of the  longitudinal and transverse polarizations   of the final baryon in the on $q^2$ for the  $\Xi_b\rightarrow \Xi_c e^-\bar{\nu}_e$ (left) and $\Xi_b\rightarrow \Xi_c \tau\bar{\nu}_\tau$ (right) transitions.}
 \label{fig:Phzx}
\end{center}
\end{figure}

\begin{figure}[!htbh]
	\begin{center}
	    \vspace{0.01cm} \centerline{\epsfxsize=12cm \epsffile{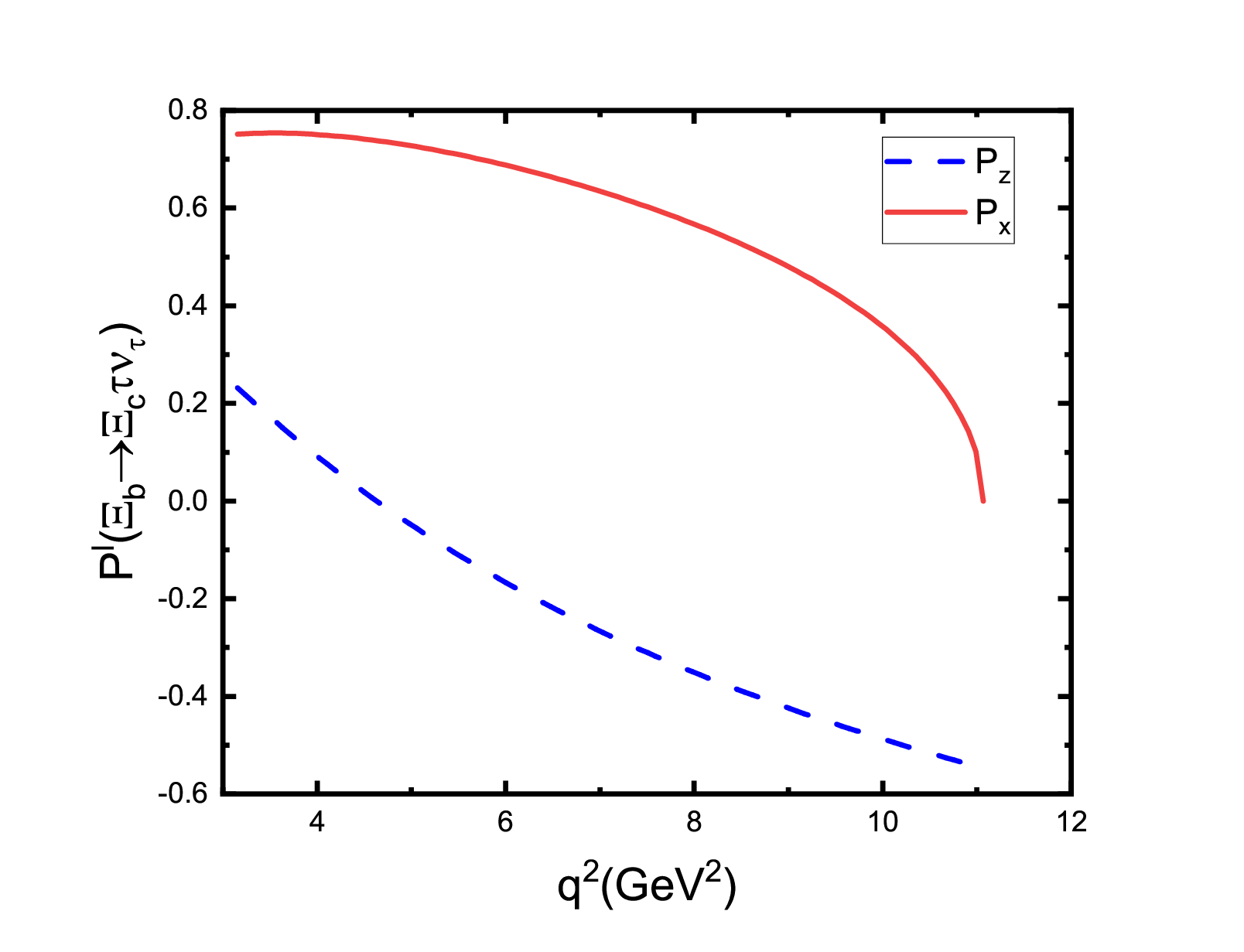}}
		\setlength{\abovecaptionskip}{1.0cm}
		\caption{$q^2$ distributions of the longitudinal and transverse polarizations of the charged lepton in the $\Xi_b\rightarrow \Xi_c \tau^-\bar{\nu}_\tau$ decay.}
		\label{fig:Plzx}
	\end{center}
\end{figure}

\item
We observe in Fig.~\ref{fig:Phzx} that the longitudinal polarizations for both the $e$ and $\tau$ modes  increase  as  $q^2$ rise, while the transverse polarizations show an opposite variation tendency. Furthermore,  the longitudinal polarizations  for both the two leptonic channels nearly overlap with each other. It is expected as the longitudinal hadronic polarization is independent of the lepton flavor.

\begin{figure}[!htbh]
	\begin{center}
	    \vspace{0.01cm} \centerline{\epsfxsize=12cm \epsffile{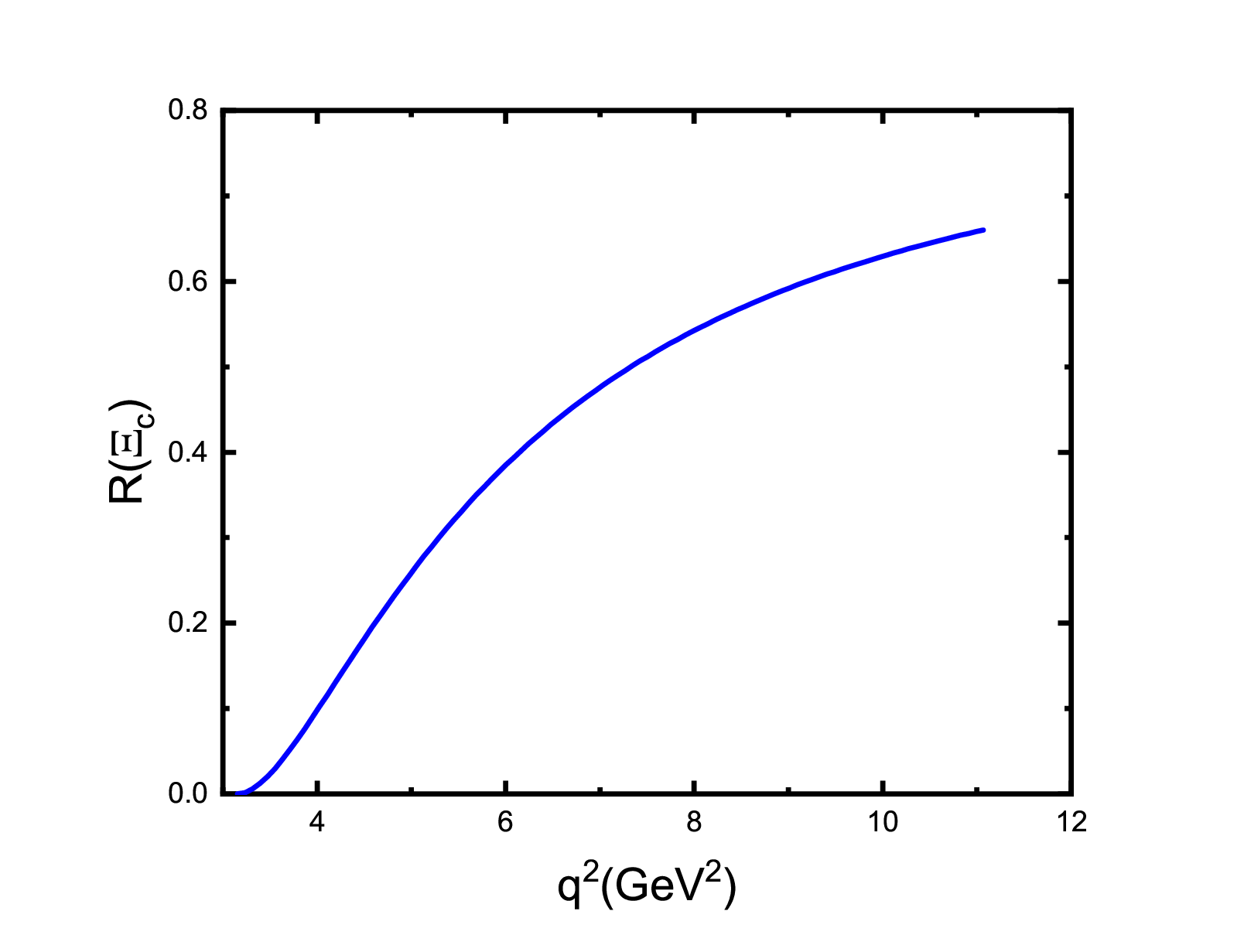}}
		\setlength{\abovecaptionskip}{1.0cm}
		\caption{$q^2$ shapes of the LFU ratio $R(q^2)$.}
		\label{fig:RR}
	\end{center}
\end{figure}

\item
As can be seen in Eq.~(\ref{eq:pl}), the longitudinal and transverse polarizations of the charged lepton are practically constant values $-1$ and 0, respectively, in the entire $q^2$ region for the $e$ mode. It is apprehensible that  the charged lepton  should be  $100\%$ polarized opposite to its momentum direction in the zero lepton mass limit. However, lepton mass effects bring in helicity flip contributions,  which can considerably change the magnitude of the polarization  and its orientation. In Fig.~\ref{fig:Plzx}, we depict the variations of the polarizations with respect to $q^2$  for the tauonic mode. It is found that both of the two curves steadily decrease with increasing $q^2$. Moreover, the transverse polarization is positive for all of the kinematic range due to the important  scalar-longitudinal interference contribution as mentioned before. Whereas for the longitudinal one, we observe a zero crossing occurs at $q^2 = 4.6\,{\rm GeV^2}$ and high, which becomes negative. The minimum value, -0.54, occurs at the  zero recoil.

\item
The rising tendency of numerical results of the LFU ratio $R(q^2)$ with the increasing of $q^2$ are shown in Fig.~\ref{fig:RR}. It is expected that the ratio should differ from unity due to the large $\tau$-lepton mass effect. It can be seen that the deviation is more pronounced at $q^2$ close to $m^2_{\tau}$.
\end{itemize}

In general, except for the longitudinal polarizations of the final baryon, we observe that the  $q^2$ distributions of the quantities for the $\tau$ mode are quite different from those for $e$ mode due to the large $\tau$-lepton mass effect. They, therefore, are crucial indicators for investigating the impact of lepton masses. The $q^2$ distributions of the above asymmetries qualitatively agree with the shapes of the SM results as reported in~\cite{Dutta:2018zqp}. Any significant deviations from these results, if observed, can be interpreted as a signal of the NP contributions.

\subsection{ Integrated observables}
\begin{table}[!htbh]
	\caption{Comparison of theoretical predictions for the
 $\Xi_b\rightarrow \Xi_c$ semileptonic decay  branching ratios and the LFU ratio.}
	\label{tab:branching}
	\begin{tabular}[t]{lcccc}
	\hline\hline
          Model                         & $\mathcal{B}_\tau$     & $\mathcal{B}_e$    &$\mathcal{R}_{\Xi_c}$\\ \hline
    This work (Exponential model)  &$(2.1^{+2.6+0.5+0.1}_{-1.0-0.2-0.2})\times 10^{-2}$     &$(7.3^{+9.3+1.7+0.2}_{-3.7-1.2-0.9})\times 10^{-2}$ &$0.295^{+0.000+0.000+0.000}_{-0.008-0.008-0.002}$ \\
    This work (QCDSR model)        &$(2.2^{+0.4+0.2+0.2}_{-0.1-0.1-0.2})\times 10^{-2}$     &$(7.6^{+1.0+0.9+1.4}_{-0.4-0.5-0.8})\times 10^{-2}$ &$0.293^{+0.001+0.003+0.003}_{-0.000-0.000-0.000}$ \\
    This work (Gegenbauer model) &$(2.5^{+0.3+0.5+0.4}_{-0.0-0.0-0.3})\times 10^{-2}$     &$(8.4^{+0.9+1.1+0.9}_{-0.0-0.0-1.1})\times 10^{-2}$ &$0.304^{+0.008+0.011+0.004}_{-0.005-0.002-0.000}$ \\
     ~\cite{Faustov:2018ahb}       &$2.0\times 10^{-2}$     &$6.15\times 10^{-2}$&$0.325$ \\
     ~\cite{Dutta:2018zqp}        &$2.35\times 10^{-2}$    &$9.22\times 10^{-2}$  &$0.255$\\
     ~\cite{Zhao:2020mod}         &$\cdots$       &$(9.02\pm 0.79)\times 10^{-2}$&$\cdots$ \\
     ~\cite{Neishabouri:2025abl}  &$(2.81^{+1.50}_{-1.15})\times 10^{-2}$
     &$(8.18^{+4.36}_{-3.34})\times 10^{-2}$      &$0.34\pm0.15$\\
		\hline\hline
	\end{tabular}
\end{table}

\begin{table}[!htbh]
	\caption{Comparison of theoretical predictions for the
 $\Xi_b\rightarrow \Xi_c$ semileptonic decay  branching fractions and LFU ratio.
 The upper and lower rows are the values for the lepton type $\tau$ or $e$, respectively.}
	\label{tab:obser}
	\begin{tabular}[t]{lccccccc}
	\hline\hline
 Model &$l$ & $\langle A_{\text{FB}}  \rangle$    & $\langle C_{\text{F}}\rangle $   & $\langle P_{z}^{h}\rangle$
   & $\langle P_{x}^{h}\rangle$ & $\langle P_{z}^{\ell}\rangle$  & $\langle P_{x}^{\ell}\rangle$  \\ \hline
This work (Exponential model)            &$\tau$      &$-0.140^{+0.001}_{-0.004}$   &$-0.098^{+0.001}_{-0.001}$  &$-0.748^{+0.002}_{-0.015}$ &$-0.170^{+0.022}_{-0.001}$ &$-0.181^{+0.020}_{-0.001}$ &$0.589^{+0.010}_{-0.001}$\\
                                           &$e$       &$0.139^{+0.001}_{-0.001}$    &$-0.623^{+0.001}_{-0.022}$  &$-0.750^{+0.002}_{-0.020}$ &$-0.476^{+0.018}_{-0.000}$ &$-1$     &0\\
This work (QCDSR model)                  &$\tau$      &$-0.141^{+0.001}_{-0.000}$   &$-0.098^{+0.001}_{-0.001}$  &$-0.750^{+0.005}_{-0.001}$ &$-0.168^{+0.009}_{-0.006}$ &$-0.178^{+0.001}_{-0.003}$ &$0.590^{+0.005}_{-0.004}$\\
                                           &$e$       &$0.139^{+0.002}_{-0.002}$    &$-0.627^{+0.002}_{-0.007}$  &$-0.753^{+0.009}_{-0.000}$ &$-0.473^{+0.002}_{-0.005}$ &$-1$     &0\\
This work (Gegenbauer model)            &$\tau$     &$-0.135^{+0.003}_{-0.001}$   &$-0.099^{+0.002}_{-0.002}$  &$-0.762^{+0.018}_{-0.004}$
                                                       &$-0.159^{+0.007}_{-0.021}$ &$-0.177^{+0.003}_{-0.019}$ &$0.599^{+0.003}_{-0.014}$\\
                                           &$e$       &$0.148^{+0.003}_{-0.003}$    &$-0.618^{+0.020}_{-0.002}$  &$-0.764^{+0.022}_{-0.004}$ &$-0.463^{+0.005}_{-0.031}$ &$-1$     &0\\
~\cite{Faustov:2018ahb} &$\tau$   &$-0.018$   &$-0.087$  &$-0.703$ &$\cdots$ &$-0.324$ &$\cdots$\\
                        &$e$      &$0.199$   &$-0.540$  &$-0.794$ &$\cdots$ &$-1$ &$\cdots$\\
~\cite{Dutta:2018zqp}   &$\tau$   &$-0.042$   &$-0.103$  &$\cdots$ &$\cdots$ &$-0.317$ &$\cdots$\\
                         &$e$      &$0.163$   &$-0.697$  &$\cdots$ &$\cdots$ &$-1$ &$\cdots$\\
		\hline\hline
	\end{tabular}
\end{table}

We now proceed to calculate the integrated quantities over the available phase space $q^2\in[m_\ell^2,(M-m)^2]$. Our predictions for the branching ratios of the $\tau$ and $e$ channels and their relative ratio  are listed in Table~\ref{tab:branching} in  comparison with the previous calculations~\cite{Faustov:2018ahb, Dutta:2018zqp, Zhao:2020mod, Neishabouri:2025abl}. The three theoretical errors of our results are the same as for form factors in Table~\ref{tab:form1}. Similar to the case of form factors, the calculated branching ratios based on the three models of LCDAs yield similar values. The predictions on the branching ratio for the $\tau$ mode are consistent with the previous estimates from Refs.~\cite{Faustov:2018ahb,Dutta:2018zqp,Zhao:2020mod, Neishabouri:2025abl}. Our numbers for the $e$ mode lie between the results from~\cite{Dutta:2018zqp,Zhao:2020mod} and~\cite{Faustov:2018ahb}. Considering the theoretical uncertainties, our results still agree with them. In general, the predicted branching ratio can reach up to the magnitude of $10^{-2}$, which is accessible at LHCb in the near future.

Since the rate of the channel involving the $\tau$ lepton  undergoes kinematic suppression due to its heavier mass, the  LFU  ratio $\mathcal{R}_{\Xi_c}$ must be smaller than one. The PQCD predictions on the ratio $\mathcal{R}_{\Xi_c}$  are not sensitive to the different models of baryonic LCDAs as shown in Table~\ref{tab:branching}. Our central value is around 0.3, which is lower than the  estimations of Refs.~\cite{Faustov:2018ahb,Neishabouri:2025abl}, but slightly higher than the value  reported in Ref.~\cite{Dutta:2018zqp}. Recently, LHCb has measured the LFU ratio for the $\Lambda_b\rightarrow\Lambda_c \ell \bar {\nu}_{\ell}$ decays and obtains $\mathcal{R}_{\Lambda_c}=0.242\pm 0.026\pm 0.040\pm 0.059$~\cite{LHCb:2022piu}, where the first uncertainty is statistical, the second is systematic, and the third is due to external branching fraction measurements. This result is slightly lower than the  most precise  SM prediction $\mathcal{R}_{\Lambda_c}=0.324\pm 0.004$ \cite{Bernlochner:2018kxh} by combining the lattice information with the measured spectrum. This deviation implies a potential violation of LFU in semileptonic $\Lambda_b\rightarrow\Lambda_c \ell \bar {\nu}_{\ell}$ decays. However, the current precision of experimental data is insufficient to justify such a judgment. A more compelling argument could be constructed by  combining $\mathcal{R}_{\Xi_c}$ and $\mathcal{R}_{\Lambda_c}$ both theoretically and experimentally. As the measurement  of $\mathcal{R}_{\Xi_c}$ is not yet available at the moment, our predictions  can give an additional hint for exploring LFU in the $b\rightarrow c \ell \nu_\ell$ transition.

In Table.~\ref{tab:obser}, we give predictions for the average values of the forward-backward asymmetry of the charged lepton $\langle A_{\text{FB}}\rangle$, the lepton-side convexity parameter, $\langle C_{\text{F}}\rangle $, the hadronic polarization components, $\langle P_{z,x}^{h}\rangle$, and the leptonic  polarization components, $\langle P_{z,x}^{\ell}\rangle$, which are calculated by individually integrating the numerators and denominators over  $q^2$ inclusive of the kinematical factor $q^2|P|(1-\frac{m_\ell^2}{q^2})^2$. Note that these quantities are less sensitive to the uncertainties in the model parameters since the errors partially cancel in the ratios of the helicity structures. All the considered theoretical errors have been added in quadrature. Similar to the prediction on the form factors, we list the results from three different LCDAs models for comparison. It is found that the predicted asymmetries  are also less sensitive to the LCDAs models apparently, and the results from different LCDAs models are consistent within the uncertainties. We notice that there are considerable changes while going from the $e$ to the $\tau$ mode because of the sizable lepton mass effect, including even a sign change in the forward-backward asymmetry parameter  $\langle A_{\text{FB}}\rangle$. We also compare our results for the integrated observables with the predictions of other theoretical approaches~\cite{Faustov:2018ahb, Dutta:2018zqp} in Table.~\ref{tab:obser}. It is found that all existing theoretical predictions for these observables agree well, which could be tested by future experimental measurements.
\section{ conclusion}\label{sec:sum}
In this work, we explore the semileptonic weak decays of the  $\Xi_b \rightarrow \Xi_c \ell^- \bar{\nu}_\ell$ in terms of the PQCD. We construct three models for the charmed baryon LCDAs from the corresponding ones for the  $b$-baryon LCDAs based on the light-front formalism and  the heavy quark symmetry, which maintain  approximate on-shell conditions of  the partonic  charm quark and  obey the parton kinetic constraints. Six independent transition form factors are calculated in the low $q^2$ region and extrapolated to the whole physical kinematical range by parametrizing the $q^2$ dependence using $z$ expansions. The results of the form factors based on the QCDSR model, Exponential model, and  Gegenbauer model of baryonic LCDAs are consistent with each other. We also discuss the behavior of these form factors with respect to $q^2$, which is expected from weak decays that the magnitude of each form factor gradually increases with respect to $q^2$. The dominant form factors are $f_1$ and  $g_1$ with very similar $q^2$  dependencies. The obtained vanishing small values of $f_{2,3}(q^2)$ and $g_{2,3}(q^2)$ over the entire $q^2$ range are in line with the expectations of the  HQET. Our results at the maximal and minimal recoil points were compared with previous theoretical calculations within different approaches, including various quark model, the QCD sum rule, and the heavy quark effective theory.

Utilizing these form factors and the helicity formalism,  we further investigate some interesting experimental  observables, such as the branching ratios, forward-backward asymmetries, lepton-side convexity parameter, as well as the final state polarizations, which serve as a highly sensitive probe of NP and of significant experimental importance. Both of the differential and integrated observables across all lepton channels are presented in Sec.~\ref{sec:results}, which could be tested with future experiments. One can find that the predicted asymmetries are apparently insensitive to the different LCDAs models, and the results from different LCDAs models are consistent within the uncertainties. It is found  that the shape of the differential distributions of these quantities for the $\tau$ mode is different from those for the $e$  mode due to the large $\tau$-lepton mass effect. These predictions may confirm  the SM or give complimentary information regarding possible new physics in heavy baryon decays. In addition, we predict  the lepton flavor violating ratio $\mathcal{R}_{\Xi_c} \approx 0.3$ that shows good consistency with existing available predictions. This value is also comparable with the counterpart in the $\Lambda_b\rightarrow\Lambda_c \ell^- \bar {\nu}_{\ell}$ decays. Our predictions can give an additional hint for exploring LFU in the $b\rightarrow c \ell \nu_\ell$ transition.

\begin{acknowledgments}
We would like to acknowledge Professor Hsiang-nan Li  for helpful discussions. This work is supported by the National Natural Science Foundation of China under Grants Nos. 12075086, 12375089 and 12435004. Z.T.Z. is  supported by  the Natural Science Foundation of Shandong Province under the Grant No. ZR2022MA035. Y.Li is also supported by the Natural Science Foundation of Shandong Province under the Grant No. ZR2022ZD26.
\end{acknowledgments}
\begin{appendix}
\section{FACTORIZATION FORMULAS}\label{sec:for}
The virtualities of internal particles in Eq.~(\ref{eq:ttt}) are given in Table~\ref{tab:wil}.
The   expressions $\Omega_i$ are presented in Table~\ref{tab:bb},
where the auxiliary functions $h_l$ are defined as~\cite{Zhang:2022iun}
\begin{eqnarray}
h_1(\textbf{b},A,B)&=&\frac{1}{16\pi^3}\int_0^1 dz \frac{\sqrt{\zeta_1}}{\sqrt{\eta_1}}
\left\{K_1(\sqrt{\zeta_1\eta_1})\theta(\eta_1)+\frac{\pi}{2}[N_1(\sqrt{-\zeta_1\eta_1})-i J_1(\sqrt{-\zeta_1\eta_1})]\theta(-\eta_1)\right\}, \nonumber\\
\zeta_1(\textbf{b})&=&|\textbf{b}|^2, \nonumber\\
\eta_1(A,B)&=&Az+B\bar{z}, \\
h_2(\textbf{b}_A,\textbf{b}_B,A,B,C)&=&\frac{1}{32\pi^3}\int_0^1\frac{dz_1dz_2}{z_1\bar{z}_1} \frac{\sqrt{\zeta_2}}{\sqrt{\eta_2}}
\left\{K_1(\sqrt{\zeta_2\eta_2})\theta(\eta_2)+\frac{\pi}{2}[N_1(\sqrt{-\zeta_2\eta_2})-i J_1(\sqrt{-\zeta_2\eta_2})]\theta(-\eta_2)\right\}, \nonumber\\
\zeta_2(\textbf{b}_A,\textbf{b}_B)&=&|\textbf{b}_A-z_1 \textbf{b}_B|^2+\frac{z_1\bar{z}_1}{z_2}|\textbf{b}_B|^2, \nonumber\\
\eta_2(A,B,C)&=&A\bar{z}_2+\frac{z_2}{z_1\bar{z}_1}[B\bar{z}_1+Cz_1],
\end{eqnarray}
with the Bessel functions $K_{0,1}$, $N_1$, and $J_1$ and the Feynman parameters $z_l$ and $\bar {z}_l=1-z_l$.

\begin{table}[!htbh]
\caption{The internal gluon $t_{A,B}$ and quark $t_{C,D}$ for each diagram in Eq.~(\ref{eq:FG}). }
\label{tab:wil}
\begin{tabular}[t]{lcccc}
\hline\hline
 $k$      & $t_A$ & $t_B$ & $t_C$          & $t_D$                 \\ \hline
$a$    & $x_3x_3'f^+M^2$      & $(1-x_1)(1-x_1')f^+M^2$     & $(1-x_1)x'_3f^+M^2$              & $(1-x_1')f^+M^2$                          \\
$b$    & $x_3x_3'f^+M^2$      & $(1-x_1)(1-x_1')f^+M^2$     & $(1-x'_1)x_3f^+M^2$              & $(1-x_1')f^+M^2$                           \\
$c$    & $x_3x_3'f^+M^2$      & $x_2x_2'f^+M^2$             & $(x_2+(1-x_2)x'_3f^+)M^2$          & $(1-x_1')f^+M^2$                          \\
$d$    & $x_3x_3'f^+M^2$      & $x_2x_2'f^+M^2$             & $(r_c^2+x_3(1-x'_2)f^+)M^2$          & $(x_3+x'_2(1-x_3)f^+)M^2$                          \\
$e$    & $x_3x_3'f^+M^2$      & $(1-x_1)(1-x_1')f^+M^2$             &  $(1-x_1)x'_3f^+M^2$          & $(r_c^2+(1-x_1)f^+)M^2$                           \\
$f$    & $x_3x_3'f^+M^2$      & $(1-x_1)(1-x_1')f^+M^2$             &  $(1-x'_1)x_3f^+M^2$          & $(r_c^2+(1-x_1)f^+)M^2$                           \\
$g$    & $x_3x_3'f^+M^2$      & $x_2x_2'f^+M^2$             &  $(r_c^2+(1-x_1)f^+)M^2$ &      $(r_c^2+(1-x'_3)x_2f^+)M^2$                    \\
\hline\hline
\end{tabular}
\end{table}

\begin{table}[H]
\footnotesize
\centering
	\caption{The expressions of  $\Omega_{k}$  in Eq.~(\ref{eq:FG}).}
	\label{tab:bb}
	\begin{tabular}[t]{lcc}
		\hline\hline
$k$  &$\Omega_{k}$\\ \hline
$a$  &$K_0(\sqrt{t_A}|\textbf{b}_2-\textbf{b}_3|)
       K_0(\sqrt{t_B}|\textbf{b}_3-\textbf{b}'_2+\textbf{b}'_3|)
       K_0(\sqrt{t_C}|\textbf{b}_2-\textbf{b}_3+\textbf{b}'_2-\textbf{b}'_3|)
       K_0(\sqrt{t_D}|\textbf{b}_3+\textbf{b}'_3|)$\\
$b$  &$K_0(\sqrt{t_A}|\textbf{b}'_2-\textbf{b}'_3|)
       K_0(\sqrt{t_B}|\textbf{b}_2|)
       K_0(\sqrt{t_C}|\textbf{b}_2-\textbf{b}_3+\textbf{b}'_2-\textbf{b}'_3|)
       K_0(\sqrt{t_D}|\textbf{b}_3+\textbf{b}'_3|)$\\	
 $c$   &$K_0(\sqrt{t_A}|\textbf{b}_3|) h_2(|\textbf{b}'_2|,|\textbf{b}_2+\textbf{b}'_2|,t_B,t_C,t_D) \delta^2(\textbf{b}_2-\textbf{b}_3+\textbf{b}'_2-\textbf{b}'_3)$\\
 $d$   &$K_0(\sqrt{t_A}|\textbf{b}_3-\textbf{b}_2-\textbf{b}'_2|)
         K_0(\sqrt{t_B}|\textbf{b}_2|)
         h_1(|\textbf{b}_2+\textbf{b}'_2|,t_C,t_D) \delta^2(\textbf{b}_2-\textbf{b}_3+\textbf{b}'_2-\textbf{b}'_3)$\\
 $e$  &$K_0(\sqrt{t_A}|\textbf{b}_2-\textbf{b}_3|)
       K_0(\sqrt{t_B}|\textbf{b}'_2|)
       K_0(\sqrt{t_C}|\textbf{b}_2-\textbf{b}_3+\textbf{b}'_2-\textbf{b}'_3|)
       K_0(\sqrt{t_D}|\textbf{b}_3+\textbf{b}'_3|)$\\
 $f$  &$K_0(\sqrt{t_A}|\textbf{b}'_2-\textbf{b}'_3|)
       K_0(\sqrt{t_B}|\textbf{b}_2-\textbf{b}_3-\textbf{b}'_3|)
       K_0(\sqrt{t_C}|\textbf{b}_2-\textbf{b}_3+\textbf{b}'_2-\textbf{b}'_3|)
       K_0(\sqrt{t_D}|\textbf{b}_3+\textbf{b}'_3|)$\\
 $g$   &$K_0(\sqrt{t_B}|\textbf{b}'_2|) h_2(|\textbf{b}_3|,|\textbf{b}_2+\textbf{b}'_2|,t_A,t_C,t_D) \delta^2(\textbf{b}_2-\textbf{b}_3+\textbf{b}'_2-\textbf{b}'_3)$\\
 \hline\hline
	\end{tabular}
\end{table}

The formulas of $H_k(x_i,x'_i)$ for the vector form factors from the Feynman diagrams $k=a-g$ are as below

\begin{eqnarray}
H^{F_1}_a(x_i,x'_i)&=&\frac{M^2}{2(f^--f^+)} ((x_2+x_3) (-f^-+f^++f^+ (f^-+r) (x'_2+x'_3)) (2 \phi _{3 s}^b (\phi _{3 a}^c-\phi _{3 s}^c)+\phi
   _2^b \phi _4^c)\nonumber\\&&+(f^--f^+) f^+ x'_3 (\phi _4^b \phi _2^c-2 \phi _{3 s}^b (\phi _{3 a}^c+\phi _{3
   s}^c))),\nonumber\\
   H^{F_2}_a(x_i,x'_i)&=&0, \nonumber\\
   H^{F_3}_a(x_i,x'_i)&=&\frac{M^2}{(f^--f^+)}f^+  (- (x_2+x_3 ) )  (x'_2+x'_3 )  (2 \phi _{3 s}^b  (\phi _{3 a}^c-\phi _{3 s}^c )+\phi _2^b \phi _4^c ),
\end{eqnarray}

\begin{eqnarray}
H^{F_1}_b(x_i,x'_i)&=&\frac{M^2}{2(f^--f^+)} ((f^--f^+) f^+ (x'_2+x'_3) (2 \phi _{3 s}^c (\phi _{3 a}^b+\phi _{3 s}^b)-\phi _2^b \phi
   _4^c)\nonumber\\&&-x_3 (-f^-+f^++f^+ (f^-+r) (x'_2+x'_3)) (2 \phi _{3 s}^c (\phi _{3 a}^b-\phi _{3 s}^b)+\phi
   _4^b \phi _2^c)),\nonumber\\
   H^{F_2}_b(x_i,x'_i)&=&0, \nonumber\\
   H^{F_3}_b(x_i,x'_i)&=&\frac{M^2}{(f^--f^+)}f^+ x_3 (x'_2+x'_3) (2 \phi _{3 s}^c (\phi _{3 a}^b-\phi _{3 s}^b)+\phi _4^b \phi _2^c),
\end{eqnarray}

\begin{eqnarray}
H^{F_1}_c(x_i,x'_i)&=&\frac{M^2}{4(f^--f^+)} ((\phi _4^c (x'_3 f^+ (f^+-f^-)+x_2 (-f^-+(x'_2+x'_3) (r+f^-) f^++f^+))\nonumber\\&& -2
   (x_2-1) \phi _2^c (f^--f^+)) \phi _2^b+2 r \phi _4^b \phi _4^c (x')_3^2 (f^+)^2+2 \phi _4^b \phi _4^c
   x'_3 (f^+)^2+2 \phi _{3 a}^b \phi _{3 a}^c x'_3 (f^+)^2 \nonumber\\&& -2 \phi _{3 s}^b \phi _{3 s}^c x'_3 (f^+)^2+2 r \phi _4^b \phi
   _4^c x'_2 x'_3 (f^+)^2+2 \phi _4^b \phi _4^c (x')_3^2 f^- (f^+)^2+2 \phi _4^b \phi _4^c x'_2 x'_3 f^-
   (f^+)^2\nonumber\\&&+2 \phi _4^b \phi _4^c f^-+4 \phi _{3 a}^b \phi _{3 a}^c f^-+4 \phi _{3 s}^b \phi _{3 s}^c f^--2 \phi _4^b \phi _4^c f^+-4 \phi _{3 a}^b
   \phi _{3 a}^c f^+-4 \phi _{3 s}^b \phi _{3 s}^c f^+-2 r \phi _4^b \phi _4^c x'_2 f^+\nonumber\\&& -2 r \phi _{3 a}^b \phi _{3 a}^c x'_2 f^++2 r \phi _{3 s}^b \phi _{3
   a}^c x'_2 f^++2 r \phi _{3 a}^b \phi _{3 s}^c x'_2 f^+-2 r \phi _{3 s}^b \phi _{3 s}^c x'_2 f^+-2 r \phi _4^b \phi _4^c x'_3 f^+-2 r \phi _{3 a}^b \phi
   _{3 a}^c x'_3 f^+\nonumber\\&&+2 r \phi _{3 s}^b \phi _{3 a}^c x'_3 f^++2 r \phi _{3 a}^b \phi _{3 s}^c x'_3 f^+-2 r \phi _{3 s}^b \phi _{3 s}^c x'_3 f^+-2 \phi _4^b
   \phi _4^c x'_2 f^- f^+-2 \phi _{3 a}^b \phi _{3 a}^c x'_2 f^- f^+\nonumber\\&&+2 \phi _{3 s}^b \phi _{3 a}^c x'_2 f^- f^++2 \phi _{3 a}^b \phi _{3 s}^c x'_2 f^- f^+-2
   \phi _{3 s}^b \phi _{3 s}^c x'_2 f^- f^+-4 \phi _4^b \phi _4^c x'_3 f^- f^+-4 \phi _{3 a}^b \phi _{3 a}^c x'_3 f^- f^+\nonumber\\&&+2 \phi _{3 s}^b \phi _{3 a}^c x'_3
   f^- f^++2 \phi _{3 a}^b \phi _{3 s}^c x'_3 f^- f^+-\phi _4^b \phi _2^c x'_3 (f^--f^+) f^+\nonumber\\&&+x_2 (\phi _2^c \phi _4^b+2 \phi _{3 a}^b \phi
   _{3 a}^c-2 \phi _{3 s}^b \phi _{3 s}^c) (-f^-+(x'_2+x'_3) (r+f^-) f^++f^+)),\nonumber\\
   H^{F_2}_c(x_i,x'_i)&=&0, \nonumber\\
   H^{F_3}_c(x_i,x'_i)&=&-\frac{M^2}{2(f^--f^+)} f^+ (x'_2+x'_3) (x_2 (2 \phi _{3 a}^b \phi _{3 a}^c-2 \phi _{3 s}^b \phi _{3 s}^c+\phi _2^b \phi _4^c+\phi _4^b \phi
   _2^c)\nonumber\\&&-2 (\phi _{3 a}^b-\phi _{3 s}^b) (\phi _{3 a}^c-\phi _{3 s}^c)+2 \phi _4^b \phi _4^c (f^+ x'_3-1)),
\end{eqnarray}

\begin{eqnarray}
H^{F_1}_d(x_i,x'_i)&=&\frac{M^2}{2(f^--f^+)} (f^+ (f^++r) (x'_1+x'_3) (-((\phi _{3 a}^b+\phi _{3 s}^b) (\phi _{3 a}^c+\phi _{3
   s}^c)+\phi _4^b \phi _4^c))+r_c (\phi _{3 s}^c ((f^-+f^++2 r) \phi _{3 a}^b \nonumber\\&&+(f^+-f^-) \phi _{3
   s}^b)+\phi _{3 a}^c ((f^+-f^-) \phi _{3 a}^b+(f^-+f^++2 r) \phi _{3 s}^b)+(x_3-1) \phi _2^b \phi _2^c
   (f^-+r)\nonumber\\&&-\phi _4^b \phi _4^c (f^++r) (f^+ x'_2-1))+x_3 (\phi _{3 a}^b (\phi _{3 a}^c (f^+
   ((f^++r) x'_1+(f^++r) x'_3+r x'_2)+f^- (f^+ x'_2-1)-r)\nonumber\\&&+\phi _{3 s}^c (f^+ ((f^++r)
   x'_1+(f^++r) x'_3-r x'_2)+f^- (1-f^+ x'_2)+r))+\phi _{3 s}^b (\phi _{3 a}^c \nonumber\\&&(f^+
   ((f^++r) x'_1+(f^++r) x'_3-r x'_2)+f^- (1-f^+ x'_2)+r)+\phi _{3 s}^c (f^+ ((f^++r)
   x'_1+(f^++r) x'_3+r x'_2)\nonumber\\&&+f^- (f^+ x'_2-1)-r))+\phi _2^b (-\phi _2^c) (f^-+r))),\nonumber\\
   H^{F_2}_d(x_i,x'_i)&=&\frac{M^2}{2(f^--f^+)^2} ((2 r_c (\phi _2^b \phi _2^c-(\phi _{3 a}^b-\phi _{3 s}^b) (\phi _{3 a}^c-\phi _{3 s}^c) (x'_2
   f^+-1))\nonumber\\&&-(\phi _4^c \phi _2^b+\phi _4^b \phi _2^c+2 \phi _{3 a}^b \phi _{3 a}^c-2 \phi _{3 s}^b \phi _{3 s}^c) x'_2
   (x'_1+x'_3) (f^+)^2) (f^-)^2\nonumber\\&&+(-2 x_3^2 \phi _2^c f^+ \phi _2^b-x_3 ((r_c \phi _4^c f^+-2 \phi _2^c
   (r+r_c (f^+-r))) \phi _2^b+r_c \phi _4^b \phi _2^c f^++2 \phi _{3 s}^b (((x'_1+x'_3)
   (f^+)^2\nonumber\\&&+x'_2 (f^+-r) f^++r) \phi _{3 a}^c+\phi _{3 s}^c (f^+ (-r_c+x'_2 (r-f^+)+(x'_1+x'_3)
   f^+)-r))\nonumber\\&&+2 \phi _{3 a}^b ((f^+ (r_c+x'_2 (r-f^+)+(x'_1+x'_3) f^+)-r) \phi _{3 a}^c+\phi
   _{3 s}^c ((x'_1+x'_3) (f^+)^2\nonumber\\&&+x'_2 (f^+-r) f^++r)))+f^+ (r_c ((\phi _4^c x'_2
   (r-f^+)-2 \phi _2^c) \phi _2^b+2 \phi _{3 a}^c (\phi _{3 a}^b (x'_2 (2 r-f^+)-2)\nonumber\\&&-r \phi _{3 s}^b x'_2)+2
   \phi _{3 s}^c (\phi _{3 s}^b (x'_2 f^+-2)-r \phi _{3 a}^b x'_2)+\phi _4^b (x'_2 (r-f^+) \phi _2^c+2 \phi _4^c
   (x'_2 f^+-1)))\nonumber\\&&-(x'_1+x'_3) ((2 \phi _4^c (x'_2 (r-f^+) f^+-r)-r \phi _2^c x'_2
   f^+) \phi _4^b+r (-\phi _4^c x'_2 f^+ \phi _2^b\nonumber\\&&-2 \phi _{3 s}^b (\phi _{3 a}^c+\phi _{3 s}^c (1-x'_2 f^+))-2 \phi _{3
   a}^b ((x'_2 f^++1) \phi _{3 a}^c+\phi _{3 s}^c))))) f^-\nonumber\\&&+f^+ (r ((2 \phi _2^c-\phi _4^c)
   \phi _2^b-\phi _4^b \phi _2^c-2 \phi _{3 a}^b \phi _{3 a}^c+2 \phi _{3 s}^b \phi _{3 s}^c) x_3^2\nonumber\\&&+(\phi _4^c \phi _2^b+\phi _4^b \phi _2^c+2
   \phi _{3 a}^b \phi _{3 a}^c-2 \phi _{3 s}^b \phi _{3 s}^c) f^+ x_3^2+r ((r_c \phi _4^c-2 \phi _2^c) \phi _2^b+r_c (\phi _2^c
   \phi _4^b\nonumber\\&&+2 (2 \phi _{3 a}^b+\phi _{3 s}^b) \phi _{3 a}^c+2 \phi _{3 a}^b \phi _{3 s}^c)+2 \phi _{3 s}^c
   (((x'_1+x'_3) f^++1) \phi _{3 a}^b+\phi _{3 s}^b ((x'_1+x'_3) f^+-1))\nonumber\\&&+2 \phi _{3 a}^c
   (((x'_1+x'_3) f^+-1) \phi _{3 a}^b+\phi _{3 s}^b ((x'_1+x'_3) f^++1))) x_3-2 r_c
   ((x_3-1) (\phi _{3 a}^b+\phi _{3 s}^b) (\phi _{3 a}^c+\phi _{3 s}^c)-\phi _4^b \phi _4^c) f^+\nonumber\\&&-2 r (\phi
   _4^c (x'_1+r_c x'_2+x'_3) \phi _4^b+(\phi _{3 a}^b+\phi _{3 s}^b) (\phi _{3 a}^c+\phi _{3 s}^c)
   (x'_1+x'_3)) f^+)), \nonumber\\
   H^{F_3}_d(x_i,x'_i)&=&\frac{M^2}{2(f^--f^+)^2} (2 f^+ x_3 (\phi _{3 a}^b (\phi _{3 a}^c ((r-f^-) x'_2+(f^+-r) x'_1+(f^+-r) x'_3)+\phi
   _{3 s}^c ((f^--r) x'_2\nonumber\\&&+(f^+-r) x'_1+(f^+-r) x'_3))+\phi _{3 s}^b (\phi _{3 a}^c
   ((f^--r) x'_2+(f^+-r) x'_1+(f^+-r) x'_3)+\phi _{3 s}^c ((r-f^-) x'_2\nonumber\\&&+(f^+-r)
   x'_1+(f^+-r) x'_3)))-(f^+)^2 x'_2 (x'_1+x'_3) ((r-f^-) (2 \phi _{3 a}^b \phi _{3
   a}^c-2 \phi _{3 s}^b \phi _{3 s}^c+\phi _2^b \phi _4^c)\nonumber\\&&+\phi _4^b (\phi _2^c (r-f^-)-2 \phi _4^c
   (r-f^+)))+r_c (f^+ x'_2 (2 \phi _{3 s}^c ((r-f^-) \phi _{3 a}^b+(f^--f^+) \phi _{3
   s}^b)\nonumber\\&&+2 \phi _{3 a}^c ((f^-+f^+-2 r) \phi _{3 a}^b+(r-f^-) \phi _{3 s}^b)+\phi _2^b \phi _4^c
   (f^+-r)-\phi _4^b (\phi _2^c-2 \phi _4^c) (r-f^+))+x_3 (2 \phi _{3 s}^c ((f^+-r) \phi _{3
   a}^b\nonumber\\&&+(f^+-f^-) \phi _{3 s}^b)+2 \phi _{3 a}^c ((f^-+f^+-2 r) \phi _{3 a}^b+(f^+-r) \phi _{3 s}^b)+\phi
   _2^b (2 \phi _2^c-\phi _4^c) (r-f^-)\nonumber\\&&+\phi _4^b \phi _2^c (f^--r)))+x_3^2 ((r-f^+) (2
   \phi _{3 a}^b \phi _{3 a}^c-2 \phi _{3 s}^b \phi _{3 s}^c+\phi _4^b \phi _2^c)+\phi _2^b (\phi _4^c (r-f^+)-2 \phi _2^c
   (r-f^-)))),
\end{eqnarray}

\begin{eqnarray}
H^{F_1}_e(x_i,x'_i)&=&\frac{M^2}{2(f^--f^+)} ((f^+)^2 (x_2+x_3) (x'_1+x'_2+x'_3) (2 \phi _{3 s}^b (\phi _{3 a}^c-\phi _{3 s}^c)+\phi
   _2^b \phi _4^c)\nonumber\\&&+f^+ (-r_c ((f^-+r) x'_3 (\phi _4^b \phi _2^c-2 \phi _{3 s}^b (\phi _{3 a}^c+\phi _{3
   s}^c))+x_2 (2 \phi _{3 s}^b (\phi _{3 a}^c-\phi _{3 s}^c)\nonumber\\&&+\phi _2^b \phi _4^c)+x_3 (2 \phi _{3 s}^b (\phi _{3
   a}^c-\phi _{3 s}^c)+\phi _2^b \phi _4^c))-f^- (x_2+x_3) x'_3 (\phi _4^b \phi _2^c\nonumber\\&&-2 \phi _{3 s}^b (\phi _{3
   a}^c+\phi _{3 s}^c))+r (x_2+x_3) (2 \phi _{3 s}^b ((x'_1+x'_2+2 x'_3) \phi _{3 a}^c-(x'_1+x'_2)
   \phi _{3 s}^c)\nonumber\\&&+\phi _2^b \phi _4^c (x'_1+x'_2+x'_3)-\phi _4^b \phi _2^c x'_3))-r (x_2+x_3) r_c (2 \phi _{3
   s}^b (\phi _{3 a}^c-\phi _{3 s}^c)+\phi _2^b \phi _4^c)),\nonumber\\
   H^{F_2}_e(x_i,x'_i)&=&\frac{M^2}{(f^--f^+)} f^+ (r_c (f^- x'_3 (\phi _4^b \phi _2^c-2 \phi _{3 s}^b (\phi _{3 a}^c+\phi _{3 s}^c))+x_2 (2 \phi _{3 s}^b (\phi
   _{3 a}^c-\phi _{3 s}^c)\nonumber\\&&+\phi _2^b \phi _4^c)+x_3 (2 \phi _{3 s}^b (\phi _{3 a}^c-\phi _{3 s}^c)+\phi _2^b \phi
   _4^c))-r (x_2+x_3) (2 \phi _{3 s}^b ((x'_1+x'_2+2 x'_3) \phi _{3 a}^c\nonumber\\&&-(x'_1+x'_2) \phi _{3
   s}^c)+\phi _2^b \phi _4^c (x'_1+x'_2+x'_3)-\phi _4^b \phi _2^c x'_3)), \nonumber\\
   H^{F_3}_e(x_i,x'_i)&=&0,
\end{eqnarray}

\begin{eqnarray}
H^{F_1}_f(x_i,x'_i)&=&\frac{M^2}{2(f^--f^+)}   (r_c  (x_3  (f^++r )  (2 \phi _{3 s}^c  (\phi _{3 a}^b-\phi _{3 s}^b )+\phi _4^b \phi _2^c )\nonumber\\&&+f^+
    (f^-+r )  (x'_2+x'_3 )  (\phi _2^b \phi _4^c-2 \phi _{3 s}^c  (\phi _{3 a}^b+\phi _{3 s}^b ) ) )\nonumber\\&&+f^+  (x_2
    (f^-+r )  (x'_2+x'_3 )  (\phi _2^b \phi _4^c-2 \phi _{3 s}^c  (\phi _{3 a}^b+\phi _{3 s}^b ) )\nonumber\\&&-x_3  (2 \phi _{3
   s}^c  (\phi _{3 a}^b  ( (f^++r ) x'_1+ (f^-+f^++2 r )  (x'_2+x'_3 ) )+\phi _{3 s}^b  ( (f^--f^+ )
    (x'_2+x'_3 )\nonumber\\&&- (f^++r ) x'_1 ) )+\phi _2^b  (-\phi _4^c )  (f^-+r )  (x'_2+x'_3 )+\phi _4^b \phi
   _2^c  (f^++r )  ) ) ),\nonumber\\
   H^{F_2}_f(x_i,x'_i)&=&\frac{M^2}{(f^--f^+)} f^+  (r  (x_2  (x'_2+x'_3 )  (2 \phi _{3 s}^c  (\phi _{3 a}^b+\phi _{3 s}^b )-\phi _2^b \phi _4^c )\nonumber\\&&+x_3  (2 \phi _{3
   s}^c  ( (x'_1+2  (x'_2+x'_3 ) ) \phi _{3 a}^b-x'_1 \phi _{3 s}^b )+\phi _2^b  (-\phi _4^c )
    (x'_2+x'_3 )+\phi _4^b \phi _2^c  ) )\nonumber\\&&-r_c  (f^-  (x'_2+x'_3 )  (\phi _2^b \phi _4^c-2 \phi
   _{3 s}^c  (\phi _{3 a}^b+\phi _{3 s}^b ) )+x_3  (2 \phi _{3 s}^c  (\phi _{3 a}^b-\phi _{3 s}^b )+\phi _4^b \phi
   _2^c ) ) ), \nonumber\\
   H^{F_3}_f(x_i,x'_i)&=&0,
\end{eqnarray}

\begin{eqnarray}
H^{F_1}_g(x_i,x'_i)&=&\frac{M^2}{4(f^--f^+)}   (-f^+  (x'_1+x'_2 )  (x_3  (f^-+r )  (2 \phi _{3 a}^b \phi _{3 a}^c-2 \phi _{3 s}^b \phi _{3 s}^c+\phi _2^b \phi
   _4^c+\phi _4^b \phi _2^c )\nonumber\\&&+2 f^+ \phi _4^b \phi _4^c  (f^++r )  )+x_2  (f^+  ( (f^++r )
   x'_3- (f^--f^+ )  (x'_1+x'_2 ) )  (2 \phi _{3 a}^b \phi _{3 a}^c\nonumber\\&&-2 \phi _{3 s}^b \phi _{3 s}^c+\phi _2^b \phi _4^c+\phi _4^b
   \phi _2^c )+2 x_3 \phi _2^b \phi _2^c  (f^-+r ) )+r_c  (x_3  (f^++r )  (2 \phi _{3 a}^c \phi _{3 s}^b\nonumber\\&&+2 \phi _{3 s}^c
    (\phi _{3 a}^b-2 \phi _{3 s}^b )+\phi _2^b \phi _4^c+\phi _4^b \phi _2^c )+f^+  (\phi _4^b  (\phi _2^c  (f^-+r ) x'_3+2
   \phi _4^c  (f^++r )  (x'_1+x'_2 ) )\nonumber\\&&- (f^-+r )  (2 \phi _{3 a}^b  ( (x'_1+x'_2 ) \phi _{3
   a}^c+  \phi _{3 s}^c )+2 \phi _{3 s}^b  (  \phi _{3 a}^c\nonumber\\&&+  \phi
   _{3 s}^c )+\phi _2^b  (-\phi _4^c ) x'_3 ) )+2 x_2  (\phi _2^b \phi _2^c  (f^-+r )- (f^++r )  (\phi _{3
   a}^b-\phi _{3 s}^b )  (\phi _{3 a}^c-\phi _{3 s}^c ) ) )\nonumber\\&&+r_c^2  (2  (f^-+f^++2 r ) \phi _{3 a}^b \phi _{3 s}^c+2 \phi
   _{3 s}^b  ( (f^-+f^++2 r ) \phi _{3 a}^c+2  (f^--f^+ ) \phi _{3 s}^c )\nonumber\\&&- (f^--f^+ )  (\phi _2^b \phi _4^c+\phi _4^b
   \phi _2^c ) )+2 x_2^2 \phi _2^b \phi _2^c  (f^-+r ) ),\nonumber\\
   H^{F_2}_g(x_i,x'_i)&=&\frac{M^2}{2(f^--f^+)}   (f^+ r  (x'_1+x'_2 )  (x_3  (2 \phi _{3 a}^b \phi _{3 a}^c-2 \phi _{3 s}^b \phi _{3 s}^c+\phi _2^b \phi _4^c+\phi _4^b \phi
   _2^c )\nonumber\\&&+2 f^+ \phi _4^b \phi _4^c   )-x_2  (f^+ r x'_3  (2 \phi _{3 a}^b \phi _{3 a}^c-2 \phi _{3 s}^b \phi _{3
   s}^c+\phi _2^b \phi _4^c+\phi _4^b \phi _2^c )\nonumber\\&&+2 r_c  (f^- \phi _2^b \phi _2^c-f^+  (\phi _{3 a}^b-\phi _{3 s}^b )  (\phi _{3
   a}^c-\phi _{3 s}^c ) )+2 r x_3 \phi _2^b \phi _2^c )\nonumber\\&&+f^+ r_c  (f^-  (2 \phi _{3 a}^b  ( (x'_1+x'_2 ) \phi _{3
   a}^c+  \phi _{3 s}^c )+2 \phi _{3 s}^b   \phi _{3 a}^c+  \phi
   _{3 s}^c )+\phi _4^b  (-\phi _2^c ) x'_3 )\nonumber\\&&-x_3  (2 \phi _{3 a}^c \phi _{3 s}^b+2 \phi _{3 s}^c  (\phi _{3 a}^b-2 \phi _{3
   s}^b )+\phi _2^b \phi _4^c+\phi _4^b \phi _2^c )+\phi _4^c  (f^- \phi _2^b  (-x'_3 )\nonumber\\&&-2 f^+ \phi _4^b
    (x'_1+x'_2 ) ) )-4 r r_c^2  (\phi _{3 a}^b \phi _{3 s}^c+\phi _{3 a}^c \phi _{3 s}^b )-2 r x_2^2 \phi _2^b \phi _2^c ), \nonumber\\
   H^{F_3}_g(x_i,x'_i)&=&0.
\end{eqnarray}
The corresponding formulas for the  axial vector ones can be obtained by the following replacement:
\begin{eqnarray}
H^{G_i}_k&=&\pm H^{F_i}_k|_{r\rightarrow -r, r_c\rightarrow -r_c},
\end{eqnarray}
where the plus and minus signs refer to  $i=1$ and $i=2,3$, respectively.

\end{appendix}


\begin{thebibliography}{99}

\bibitem{Bernlochner:2021vlv}
F.~U.~Bernlochner, M.~F.~Sevilla, D.~J.~Robinson and G.~Wormser,
Semitauonic b-hadron decays: A lepton flavor universality laboratory,
Rev. Mod. Phys. \textbf{94},   015003 (2022).

\bibitem{Bifani:2018zmi}
S.~Bifani, S.~Descotes-Genon, A.~Romero Vidal and M.~H.~Schune,
Review of Lepton Universality tests in $B$ decays,
J. Phys. G \textbf{46},   023001 (2019).


\bibitem{BaBar:2012obs}
J.~P.~Lees \textit{et al.} [BaBar],
Evidence for an excess of $\bar{B} \to D^{(*)} \tau^-\bar{\nu}_\tau$ decays,
Phys. Rev. Lett. \textbf{109}, 101802 (2012).
\bibitem{BaBar:2013mob}
J.~P.~Lees \textit{et al.} [BaBar],
Measurement of an Excess of $\bar{B} \to D^{(*)}\tau^- \bar{\nu}_\tau$ Decays and Implications for Charged Higgs Bosons,
Phys. Rev. D \textbf{88},   072012 (2013).
\bibitem{Belle:2016ure}
Y.~Sato \textit{et al.} [Belle],
Measurement of the branching ratio of $\bar{B}^0 \rightarrow D^{*+} \tau^- \bar{\nu}_{\tau}$ relative to $\bar{B}^0 \rightarrow D^{*+} \ell^- \bar{\nu}_{\ell}$ decays with a semileptonic tagging method,
Phys. Rev. D \textbf{94},   072007 (2016).
\bibitem{Belle:2015qfa}
M.~Huschle \textit{et al.} [Belle],
Measurement of the branching ratio of $\bar{B} \to D^{(\ast)} \tau^- \bar{\nu}_\tau$ relative to $\bar{B} \to D^{(\ast)} \ell^- \bar{\nu}_\ell$ decays with hadronic tagging at Belle,
Phys. Rev. D \textbf{92},  072014 (2015).
\bibitem{Belle:2019rba}
G.~Caria \textit{et al.} [Belle],
Measurement of $\mathcal{R}(D)$ and $\mathcal{R}(D^*)$ with a semileptonic tagging method,
Phys. Rev. Lett. \textbf{124},   161803 (2020).
\bibitem{Belle:2017ilt}
S.~Hirose \textit{et al.} [Belle],
Measurement of the $\tau$ lepton polarization and $R(D^*)$ in the decay $\bar{B} \rightarrow D^* \tau^- \bar{\nu}_\tau$ with one-prong hadronic $\tau$ decays at Belle,
Phys. Rev. D \textbf{97},  012004 (2018).
\bibitem{LHCb:2023uiv}
R.~Aaij \textit{et al.} [LHCb],
Test of lepton flavor universality using $B^0 \rightarrow D^* \tau^+ \nu_{\tau}$ decays with hadronic $\tau$ channels,
Phys. Rev. D \textbf{108},  012018 (2023)
[erratum: Phys. Rev. D \textbf{109},  119902 (2024)].
\bibitem{LHCb:2023zxo}
R.~Aaij \textit{et al.} [LHCb],
Measurement of the ratios of branching fractions $\mathcal{R}(D^{*})$ and $\mathcal{R}(D^{0})$,
Phys. Rev. Lett. \textbf{131}, 111802 (2023).
\bibitem{LHCb:2024jll}
R.~Aaij \textit{et al.} [LHCb],
Measurement of the branching fraction ratios $R(D^{+})$ and $R(D^{*+})$ using muonic $\tau$ decays,
[arXiv:2406.03387 [hep-ex]].

\bibitem{Bigi:2016mdz}
D.~Bigi and P.~Gambino,
Revisiting $B\to D \ell \nu$,
Phys. Rev. D \textbf{94},   094008 (2016).
\bibitem{Gambino:2019sif}
P.~Gambino, M.~Jung and S.~Schacht,
The $V_{cb}$ puzzle: An update,
Phys. Lett. B \textbf{795}, 386 (2019).
\bibitem{Bordone:2019vic}
M.~Bordone, M.~Jung and D.~van Dyk,
Theory determination of $\bar{B}\to D^{(*)}\ell^-\bar\nu$ form factors at $\mathcal{O}(1/m_c^2)$,
Eur. Phys. J. C \textbf{80},  74 (2020).
\bibitem{Bernlochner:2017jka}
F.~U.~Bernlochner, Z.~Ligeti, M.~Papucci and D.~J.~Robinson,
Combined analysis of semileptonic $B$ decays to $D$ and $D^*$: $R(D^{(*)})$, $|V_{cb}|$, and new physics,
Phys. Rev. D \textbf{95},   115008 (2017)
[erratum: Phys. Rev. D \textbf{97},  059902 (2018)].
\bibitem{Jaiswal:2017rve}
S.~Jaiswal, S.~Nandi and S.~K.~Patra,
Extraction of $|V_{cb}|$ from $B\to D^{(*)}\ell\nu_\ell$ and the Standard Model predictions of $R(D^{(*)})$,
JHEP \textbf{12}, 060 (2017).
\bibitem{Martinelli:2021onb}
G.~Martinelli, S.~Simula and L.~Vittorio,
$\vert V_{cb} \vert$ and $R(D)^{(*)}$) using lattice QCD and unitarity,
Phys. Rev. D \textbf{105},   034503 (2022).
\bibitem{Bigi:2017jbd}
D.~Bigi, P.~Gambino and S.~Schacht,
$R(D^*)$, $|V_{cb}|$, and the Heavy Quark Symmetry relations between form factors,
JHEP \textbf{11}, 061 (2017).
\bibitem{FermilabLattice:2021cdg}
A.~Bazavov \textit{et al.} [Fermilab Lattice, MILC, Fermilab Lattice and MILC],
Semileptonic form factors for $B\rightarrow D^*\ell \nu $ at nonzero recoil from $2+1$-flavor lattice QCD: Fermilab Lattice~and~MILC~Collaborations,
Eur. Phys. J. C \textbf{82},  1141 (2022)
[erratum: Eur. Phys. J. C \textbf{83}, no.1, 21 (2023)].

\bibitem{HFLAV:2022esi}
Y.~S.~Amhis \textit{et al.} [HFLAV],
Averages of b-hadron, c-hadron, and \ensuremath{\tau}-lepton properties as of 2021,
Phys. Rev. D \textbf{107},   052008 (2023)
\bibitem{LHCb:2017vlu}
R.~Aaij \textit{et al.} [LHCb],
Measurement of the ratio of branching fractions $\mathcal{B}(B_c^+\,\to\,J/\psi\tau^+\nu_\tau)$/$\mathcal{B}(B_c^+\,\to\,J/\psi\mu^+\nu_\mu)$,
Phys. Rev. Lett. \textbf{120},  121801 (2018).

\bibitem{Rui:2016opu}
Z.~Rui, H.~Li, G.~x.~Wang and Y.~Xiao,
Semileptonic decays of $B_c$ meson to S-wave charmonium states in the perturbative QCD approach,
Eur. Phys. J. C \textbf{76},  564 (2016).
\bibitem{Qiao:2012vt}
C.~F.~Qiao and R.~L.~Zhu,
Estimation of semileptonic decays of $B_c$ meson to S-wave charmonia with nonrelativistic QCD,
Phys. Rev. D \textbf{87},   014009 (2013).
\bibitem{Harrison:2020nrv}
J.~Harrison \textit{et al.} [LATTICE-HPQCD],
$R(J/\psi)$ and $B_c^- \rightarrow J/\psi \ell^-\bar{\nu}_\ell$ Lepton Flavor Universality Violating Observables from Lattice QCD,
Phys. Rev. Lett. \textbf{125},   222003 (2020).
\bibitem{Dutta:2017xmj}
R.~Dutta and A.~Bhol,
$B_c \to (J/\psi,\,\eta_c)\tau\nu$ semileptonic decays within the standard model and beyond,
Phys. Rev. D \textbf{96},  076001 (2017).


\bibitem{Li:2018lxi}
Y.~Li and C.~D.~L\"u,
Recent Anomalies in B Physics,
Sci. Bull. \textbf{63}, 267 (2018).
\bibitem{Pich:2019pzg}
A.~Pich,
Flavour Anomalies,
PoS \textbf{LHCP2019}, 078 (2019).

\bibitem{Iguro:2024hyk}
S.~Iguro, T.~Kitahara and R.~Watanabe,
Global fit to $b \rightarrow c \tau\nu_\tau$ anomalies as of Spring 2024,
Phys. Rev. D \textbf{110},   7 (2024).

\bibitem{LHCb:2022piu}
R.~Aaij \textit{et al.} [LHCb],
Observation of the decay $ \Lambda_b^0\rightarrow \Lambda_c^+\tau^-\overline{\nu}_{\tau}$,
Phys. Rev. Lett. \textbf{128},   191803 (2022).




\bibitem{Detmold:2015aaa}
W.~Detmold, C.~Lehner and S.~Meinel,
$\Lambda_b \to p \ell^- \bar{\nu}_\ell$ and $\Lambda_b \to \Lambda_c \ell^- \bar{\nu}_\ell$ form factors from lattice QCD with relativistic heavy quarks,
Phys. Rev. D \textbf{92},   034503 (2015).

\bibitem{Bernlochner:2018kxh}
F.~U.~Bernlochner, Z.~Ligeti, D.~J.~Robinson and W.~L.~Sutcliffe,
New predictions for $\Lambda_b\to\Lambda_c$ semileptonic decays and tests of heavy quark symmetry,
Phys. Rev. Lett. \textbf{121},  202001 (2018).


\bibitem{Bernlochner:2018bfn}
F.~U.~Bernlochner, Z.~Ligeti, D.~J.~Robinson and W.~L.~Sutcliffe,
Precise predictions for $\Lambda_b \to \Lambda_c$ semileptonic decays,
Phys. Rev. D \textbf{99},   055008 (2019).

\bibitem{Fedele:2022iib}
M.~Fedele, M.~Blanke, A.~Crivellin, S.~Iguro, T.~Kitahara, U.~Nierste and R.~Watanabe,
Impact of $\Lambda_b\rightarrow \Lambda_c \tau\nu$  measurement on new physics in $b\rightarrow c \tau\nu$  transitions,
Phys. Rev. D \textbf{107},   055005 (2023).

\bibitem{Duan:2024ayo}
W.~F.~Duan, S.~Iguro, X.~Q.~Li, R.~Watanabe and Y.~D.~Yang,
Sum rules for semi-leptonic $b \to c$ and $b \to u$ decays: accuracy checks and implications,
[arXiv:2410.21384 [hep-ph]].
\bibitem{Endo:2025fke}
M.~Endo, S.~Iguro, S.~Mishima and R.~Watanabe,
Heavy quark symmetry behind $b \to c$ semileptonic sum rule,
[arXiv:2501.09382 [hep-ph]].

\bibitem{LHCb:2023ngz}
R.~Aaij \textit{et al.} [LHCb],
Observation of $\Xi_b^0 \rightarrow \Xi_c^+ D_s^-$ and $\Xi_b^- \rightarrow \Xi_c^0 D_s^-$ decays,
Eur. Phys. J. C \textbf{84},  237 (2024).

\bibitem{Cheng:1996cs}
H.~Y.~Cheng,
Nonleptonic weak decays of bottom baryons,
Phys. Rev. D \textbf{56}, 2799 (1997)
[erratum: Phys. Rev. D \textbf{99}, 079901 (2019)].

\bibitem{Ebert:2006rp}
  D.~Ebert, R.~N.~Faustov and V.~O.~Galkin,
  Semileptonic decays of heavy baryons in the relativistic quark model,
  Phys.\ Rev.\ D {\bf 73}, 094002 (2006).

\bibitem{Singleton:1990ye}
  R.~L.~Singleton,
  Semileptonic baryon decays with a heavy quark,
  Phys.\ Rev.\ D {\bf 43}, 2939 (1991).

\bibitem{Cheng:1995fe}
  H.~Y.~Cheng and B.~Tseng,
  1/M corrections to baryonic form-factors in the quark model,
  Phys.\ Rev.\ D {\bf 53}, 1457 (1996)
  Erratum: [Phys.\ Rev.\ D {\bf 55}, 1697 (1997)].

\bibitem{Ivanov:1996fj}
  M.~A.~Ivanov, V.~E.~Lyubovitskij, J.~G.~Korner and P.~Kroll,
  Heavy baryon transitions in a relativistic three quark model,
  Phys.\ Rev.\ D {\bf 56}, 348 (1997).

\bibitem{Ivanov:1998ya}
  M.~A.~Ivanov, J.~G.~Korner, V.~E.~Lyubovitskij and A.~G.~Rusetsky,
  Charm and bottom baryon decays in the Bethe-Salpeter approach: Heavy to heavy semileptonic transitions,
  Phys.\ Rev.\ D {\bf 59}, 074016 (1999).

\bibitem{Cardarelli:1998tq}
  F.~Cardarelli and S.~Simula,
  Analysis of the $\Lambda_b\rightarrow \Lambda_c \ell \nu_\ell$  decay within a light front constituent quark model,
  Phys.\ Rev.\ D {\bf 60}, 074018 (1999).

\bibitem{Albertus:2004wj}
  C.~Albertus, E.~Hernandez and J.~Nieves,
  Nonrelativistic constituent quark model and HQET combined study of semileptonic decays of Lambda(b) and Xi(b) baryons,
  Phys.\ Rev.\ D {\bf 71}, 014012 (2005).

\bibitem{Korner:1994nh}
  J.~G.~Korner, M.~Kramer and D.~Pirjol,
  Heavy baryons,
  Prog.\ Part.\ Nucl.\ Phys.\  {\bf 33}, 787 (1994).

\bibitem{Zhang:2019xdm}
J.~Zhang, X.~An, R.~Sun and J.~Su,
Probing new physics in semileptonic $\Xi _{b}\rightarrow \Lambda (\Xi _{c})\tau ^{-}\bar{\nu }_{\tau }$ decays,
Eur. Phys. J. C \textbf{79},  863 (2019).
\bibitem{Ke:2024aux}
H.~W.~Ke, G.~Y.~Fang and Y.~L.~Shi,
Study on the mixing of $\Xi_c$  and $\Xi'_c$ by the transition $\Xi_b\rightarrow \Xi^{(')}_c$,
Phys. Rev. D \textbf{109}, 073006 (2024).








\bibitem{Dutta:2018zqp}
R.~Dutta,
Phenomenology of $\Xi_b \to \Xi_c\,\tau\,\nu$ decays,
Phys. Rev. D \textbf{97},  073004 (2018).

\bibitem{Neishabouri:2025abl}
Z.~Neishabouri and K.~Azizi,
Investigation of the semileptonic decays $\Xi^{(')}_{b}\rightarrow \Xi^{(')}_{c}{\ell}\bar\nu_{\ell}$,
[arXiv:2503.12390 [hep-ph]].

\bibitem{Keum:2000wi}
Y.~Y.~Keum, H.~N.~Li and A.~I.~Sanda,
Penguin enhancement and $B \to K \pi$ decays in perturbative QCD,
Phys. Rev. D \textbf{63}, 054008 (2001).

\bibitem{Lu:2000em}
C.~D.~Lu, K.~Ukai and M.~Z.~Yang,
Branching ratio and CP violation of $B \rightarrow \pi \pi$ decays in perturbative QCD approach,
Phys. Rev. D \textbf{63}, 074009 (2001).


\bibitem{Kurimoto:2001zj}
T.~Kurimoto, H.~n.~Li, and A.~I.~Sanda,
Leading power contributions to $B\rightarrow \pi, \rho$ transition form-factors,
Phys. Rev. D \textbf{65} (2001) 014007 [hep-ph/0105003].

\bibitem{Lu:2002ny}
C.~D.~Lu and M.~Z.~Yang,
B to light meson transition form-factors calculated in perturbative QCD approach,
Eur. Phys. J. C \textbf{28}  (2003) 515 [hep-ph/0212373].

\bibitem{Ali:2007ff}
A.~Ali, G.~Kramer, Y.~Li, C.~D.~Lu, Y.~L.~Shen, W.~Wang, and Y.~M.~Wang,
Charmless non-leptonic $B_s$ decays to $PP$, $PV$ and $VV$ final states in the pQCD approach,
Phys. Rev. D \textbf{76} (2007) 074018 [hep-ph/0703162].

\bibitem{Wang:2010ni}
W.~Wang,
B to tensor meson form factors in the perturbative QCD approach,
Phys. Rev. D \textbf{83} (2011) 014008 [arXiv: 1008.5326].

\bibitem{Rui:2021kbn}
Z.~Rui, Y.~Li, and H.~n.~Li,
Four-body decays $B_{(s)} \rightarrow (K\pi)_{S/P} (K\pi)_{S/P}$ in the perturbative QCD approach,
JHEP 05 (2021) 082.

\bibitem{Chai:2022ptk}
J.~Chai, S.~Cheng, Y.~h.~Ju, D.~C.~Yan, C.~D.~L\"u, and Z.~J.~Xiao,
Charmless two-body $B$ meson decays in the perturbative QCD factorization approach,
Chin. Phys. C \textbf{46} (2022) 123103.

\bibitem{Rui:2023fiz}
Z.~Rui and Z.~T.~Zou,
Charmonium decays of beauty baryons in the perturbative QCD approach,
Phys. Rev. D \textbf{109}  (2024) 033013.

\bibitem{plb665197}
P.~Ball, V.~M.~Braun, and E.~Gardi,
Distribution amplitudes of the $\Lambda_b$ baryon in QCD,
Phys. Lett. B \textbf{665} (2008) 197.


\bibitem{Ali:2012zza}
A.~Ali, C.~Hambrock, and A.~Y.~Parkhomenko,
Light-cone wave functions of heavy baryons,
Theor. Math. Phys. \textbf{170} (2012)  2.






\bibitem{epjc732302}
A.~Ali, C.~Hambrock, A.~Y.~Parkhomenko, and W.~Wang,
Light-Cone distribution amplitudes of the ground state bottom baryons in HQET,
Eur. Phys. J. C \textbf{73} (2013) 2302.

\bibitem{plb738334}
V.~M.~Braun, S.~E.~Derkachov, and A.~N.~Manashov,
Integrability of the evolution equations for heavy-light baryon distribution amplitudes,
Phys. Lett. B \textbf{738}  (2014) 334.

\bibitem{jhep022016179}
Y.~M.~Wang and Y.~L.~Shen,
Perturbative corrections to $\Lambda_b \rightarrow \Lambda$ form factors from QCD Light-Cone Sum Rules,
 JHEP 02 (2016) 179.

\bibitem{jhep112013191}
G.~Bell, T.~Feldmann, Y.~M.~Wang, and M.~W.~Y.~Yip,
Light-cone distribution amplitudes for heavy-quark hadrons,
JHEP 11 (2013) 191.

\bibitem{Han:2024min}
X.~Y.~Han, J.~Hua, X.~Ji, C.~D.~L\"u, W.~Wang, J.~Xu, Q.~A.~Zhang and S.~Zhao,
A new method to access heavy meson lightcone distribution amplitudes from first-principle,
[arXiv:2403.17492 [hep-ph]].

\bibitem{Han:2024yun}
X.~Y.~Han, J.~Hua, X.~Ji, C.~D.~L\"u, A.~Sch\"afer, Y.~Su, W.~Wang, J.~Xu, Y.~Yang and J.~H.~Zhang, \textit{et al.}
Calculation of heavy meson light-cone distribution amplitudes from lattice QCD,
[arXiv:2410.18654 [hep-lat]].
\bibitem{Wang:2024wwa}
W.~Wang, J.~Xu, Q.~A.~Zhang and S.~Zhao,
Mass renormalization group of heavy meson light-cone distribution amplitude in QCD,
[arXiv:2411.07101 [hep-ph]].
\bibitem{Schlumpf:1992ce}
F.~Schlumpf,
Relativistic constituent quark model for baryons,
[arXiv:hep-ph/9211255 [hep-ph]].


\bibitem{Zhang:2022iun}
C.~Q.~Zhang, J.~M.~Li, M.~K.~Jia and Z.~Rui,
Nonleptonic two-body decays of $\Lambda_b\rightarrow \Lambda_c \pi, \Lambda_c K$ in the perturbative QCD approach,
Phys. Rev. D \textbf{105},  073005 (2022).

\bibitem{Kurimoto:2002sb}
T.~Kurimoto, H.~n.~Li and A.~I.~Sanda,
 $B\rightarrow D^{(*)}$ form-factors in perturbative QCD,
Phys. Rev. D \textbf{67}, 054028 (2003).


\bibitem{Shih:1999yh}
H.~H.~Shih, S.~C.~Lee and H.~n.~Li,
Applicability of perturbative QCD to $\Lambda_b\rightarrow \Lambda_c $  decays,
Phys. Rev. D \textbf{61}, 114002 (2000).

\bibitem{Wang:2010fq}
Z.~G.~Wang,
Analysis of the ${\frac{1}{2}}^{\pm}$ antitriplet heavy baryon states with QCD sum rules,
Eur. Phys. J. C \textbf{68} (2010) 479 [arXiv: 1001.1652].

\bibitem{Li:2021qod}
Y.~S.~Li, X.~Liu and F.~S.~Yu,
Revisiting semileptonic decays of $\Lambda_{b(c)}$ supported by baryon spectroscopy,
Phys. Rev. D \textbf{104},  013005 (2021).

\bibitem{prd59094014}
H.~H.~Shih, S.~C.~Lee, and H.~n.~Li,
The $\Lambda_b \to p l \bar{\nu}$ decay in perturbative QCD,
Phys. Rev. D \textbf{59} (1999) 094014 [hep-ph/9810515].

\bibitem{prd80034011}
C.~D.~Lu, Y.~M.~Wang, H.~Zou, A.~Ali, and G.~Kramer,
Anatomy of the pQCD approach to the baryonic decays $\Lambda_b\to p\pi, pK$,
Phys. Rev. D \textbf{80} (2009) 034011.

\bibitem{Liu:2023kxr}
X.~Liu,
Bc-meson decays into J/\ensuremath{\psi} plus a light meson in the improved perturbative QCD formalism,
Phys. Rev. D \textbf{108}  (2023) 096006.

\bibitem{Liu:2020upy}
X.~Liu, H.~n.~Li and Z.~J.~Xiao,
Next-to-leading-logarithm $k_T$ resummation for $B_c\to J/\psi$ decays,
Phys. Lett. B \textbf{811}, 135892 (2020).

 \bibitem{pdg2024}
Particle Data collaboration, Review of Particle Physics, Phys. Rev. D \textbf{110}  (2024) 030001.

\bibitem{Rui:2024xgc}
Z.~Rui, Z.~T.~Zou and Y.~Li,
Higher twist corrections to doubly-charmed baryonic $B$ decays,
JHEP \textbf{12}, 159 (2024).






\bibitem{Ray:2018hrx}
A.~Ray, S.~Sahoo and R.~Mohanta,
Probing new physics in semileptonic $\Lambda_b$ decays,
Phys. Rev. D \textbf{99},  015015 (2019).


\bibitem{Korner:1989qb}
J.~G.~Korner and G.~A.~Schuler,
Exclusive Semileptonic Heavy Meson Decays Including Lepton Mass Effects,
Z. Phys. C \textbf{46}, 93 (1990).

\bibitem{Gutsche:2013pp}
T.~Gutsche, M.~A.~Ivanov, J.~G.~Korner, V.~E.~Lyubovitskij and P.~Santorelli,
Rare baryon decays $\Lambda_b \to \Lambda {l^{+}l^{-}} (l=e, \mu, \tau)$ and $\Lambda_b \to \Lambda\gamma$ : differential and total rates, lepton- and hadron-side forward-backward asymmetries,
Phys. Rev. D \textbf{87}, 074031 (2013).

\bibitem{Datta:2017aue}
A.~Datta, S.~Kamali, S.~Meinel and A.~Rashed,
Phenomenology of $ {\Lambda}_b\to {\Lambda}_c\tau {\overline{\nu}}_{\tau } $ using lattice QCD calculations,
JHEP \textbf{08}, 131 (2017).

\bibitem{Kadeer:2005aq}
A.~Kadeer, J.~G.~Korner and U.~Moosbrugger,
Helicity analysis of semileptonic hyperon decays including lepton mass effects,
Eur. Phys. J. C \textbf{59}, 27  (2009).

\bibitem{Faustov:2016pal}
R.~N.~Faustov and V.~O.~Galkin,
Semileptonic decays of $\Lambda_b$ baryons in the relativistic quark model,
Phys. Rev. D \textbf{94},   073008 (2016).



\bibitem{Gutsche:2015mxa}
T.~Gutsche, M.~A.~Ivanov, J.~G.~K\"orner, V.~E.~Lyubovitskij, P.~Santorelli and N.~Habyl,
Semileptonic decay $\Lambda_b \to \Lambda_c + \tau^- + \bar{\nu_\tau}$ in the covariant confined quark model,
Phys. Rev. D \textbf{91},   074001 (2015)
[erratum: Phys. Rev. D \textbf{91}, 119907 (2015)].


\bibitem{Bialas:1992ny}
P.~Bialas, J.~G.~Korner, M.~Kramer and K.~Zalewski,
Joint angular decay distributions in exclusive weak decays of heavy mesons and baryons,
Z. Phys. C \textbf{57}, 115  (1993).


\bibitem{Albertus:2005ud}
C.~Albertus, J.~M.~Flynn, E.~Hernandez, J.~Nieves and J.~M.~Verde-Velasco,
Semileptonic $B\rightarrow \pi$ decays from an Omnes improved nonrelativistic constituent quark model,
Phys. Rev. D \textbf{72}, 033002 (2005).

\bibitem{Li:2022hcn}
Y.~S.~Li and X.~Liu,
Investigating the transition form factors of $\Lambda_b\rightarrow \Lambda_c(2625)$  and $\Xi_b\rightarrow \Xi_c(2815)$ and the corresponding weak decays with support from baryon spectroscopy,
Phys. Rev. D \textbf{107},   033005 (2023).
\bibitem{Faustov:2018ahb}
R.~N.~Faustov and V.~O.~Galkin,
Relativistic description of the $\Xi_b$ baryon semileptonic decays,
Phys. Rev. D \textbf{98}, 093006 (2018).

\bibitem{Sasaki:2008ha}
S.~Sasaki and T.~Yamazaki,
Lattice study of flavor SU(3) breaking in hyperon beta decay,
Phys. Rev. D \textbf{79}, 074508 (2009).
\bibitem{Mannel:1990vg}
T.~Mannel, W.~Roberts and Z.~Ryzak,
Baryons in the heavy quark effective theory,
Nucl. Phys. B \textbf{355}, 38  (1991).


\bibitem{Li:2021kfb}
Y.~S.~Li and X.~Liu,
Restudy of the color-allowed two-body nonleptonic decays of bottom baryons $\Xi_b$ and $\Omega_b$ supported by hadron spectroscopy,
Phys. Rev. D \textbf{105}, 013003 (2022).

\bibitem{Chua:2019yqh}
C.~K.~Chua,
Color-allowed bottom baryon to $s$-wave and $p$-wave charmed baryon nonleptonic decays,
Phys. Rev. D \textbf{100},  034025 (2019).


\bibitem{Zhao:2018zcb}
Z.~X.~Zhao,
Weak decays of heavy baryons in the light-front approach,
Chin. Phys. C \textbf{42},  093101 (2018).



\bibitem{Zhao:2020mod}
Z.~X.~Zhao, R.~H.~Li, Y.~L.~Shen, Y.~J.~Shi and Y.~S.~Yang,
The semi-leptonic form factors of $\Lambda_{b}\to\Lambda_{c}$ and $\Xi_{b}\to\Xi_{c}$ in QCD sum rules,
Eur. Phys. J. C \textbf{80},   1181 (2020).



\bibitem{Manohar:2000dt}
A.~V.~Manohar and M.~B.~Wise,
Heavy quark physics,
Camb. Monogr. Part. Phys. Nucl. Phys. Cosmol. \textbf{10}, 1-191 (2000).








































\end{thebibliography}
\end{document}